%% file: mirage-perspective-arxiv.tex
\PassOptionsToPackage{numbers,sort&compress}{natbib}
\documentclass[11pt]{article}
\usepackage[utf8]{inputenc}
\usepackage[T1]{fontenc}
\usepackage{graphicx}
\usepackage{natbib}
\usepackage{amssymb}
\usepackage{amsmath}
\usepackage[labelfont=bf]{caption} %
\usepackage{float}
\usepackage{bm}
\usepackage{xcolor}
\usepackage{booktabs}
\usepackage{tabularx}
\usepackage{makecell}
\usepackage{geometry}
\usepackage{lscape}
\usepackage{parskip}
\usepackage{wrapfig}
\usepackage{float}
\usepackage{soul}
\usepackage{comment}
\usepackage{longtable}
\usepackage{array}
\usepackage{tcolorbox}
\usepackage{multicol}
\usepackage{authblk}

\usepackage{url}  %

\usepackage[colorlinks=true, linkcolor=blue, citecolor=blue, urlcolor=blue]{hyperref}

\title{Materials Behavior as Mechanism Ensembles: A Probabilistic Framework for Emergent Behaviors}
\author[1]{Brad L. Boyce}
\author[2]{Mitchell A. Wood}
\author[3]{Krishna Garikipati}
\author[2]{Andreas E. Robertson}
\author[4]{Jeffrey Larson}
\author[5]{Ishan Srivastava}
\author[6]{Bert Debusschere}
\author[1]{Saaketh Desai}
\author[1]{Prasad Iyer}
\author[6]{Pieterjan Robbe}
\author[4]{Mathew Cherukara}
\author[4]{Todd Munson}
\author[4]{Ming Du}
\author[4]{Trupti Mohanty}
\author[7]{David J. Gardner}
\author[8]{Laurent Capolungo}
\author[3]{Benjamin A. Jasperson}
\author[3]{Jingye Tan}
\author[1]{R\'emi Dingreville\thanks{Corresponding author: \texttt{rdingre@sandia.gov}}}

\affil[1]{Center for Integrated Nanotechnologies, Sandia National Laboratories, Albuquerque, NM, USA}
\affil[2]{Center for Computing Research, Sandia National Laboratories, Albuquerque, NM, USA}
\affil[3]{Department of Aerospace and Mechanical Engineering, Univ.~of Southern California, Los Angeles, CA, USA}
\affil[4]{Argonne National Laboratory, Lemont, IL, USA}
\affil[5]{Lawrence Berkeley National Laboratory, Berkeley, CA, USA}
\affil[6]{Sandia National Laboratories, Livermore, CA, USA}
\affil[7]{Lawrence Livermore National Laboratory, Livermore, CA, USA}
\affil[8]{Los Alamos National Laboratory, Los Alamos, NM, USA}

\date{}

\begin{document}

\maketitle

\begin{abstract}
Materials behavior is often treated as a deterministic mapping from structure to properties, yet many important phenomena emerge from the conditional activation of multiple mechanisms across scales.
This is especially evident in fatigue of metals, where crack growth is typically modeled as monotonic and irreversible process, despite evidence that local microstructure, loading history, and competing unit processes can shift the balance among propagation, arrest, and self-healing.
Here we present a probabilistic framework that describes materials behavior as an ensemble of constituent mechanisms whose activation, interaction, and evolution determine emergent outcomes.
The framework connects mechanism activation, state evolution, and macroscopic observables in a probabilistic way.
In the case of fatigue crack propagation, it reframes damage tolerance as an inference problem over mechanism competition and provides a basis for integrating multiscale simulation, multimodal characterization, and machine learning.
The same logic extends to other physical and chemical
systems suggesting a portable framework for any system in which emergent behavior reflects mechanism competition under changing conditions.
The broader ambition of this perspective review is a shift from correlating structure and performance after the fact to identifying, in advance, the conditions that make desired emergent behavior probable.

\end{abstract}

\begin{figure}[h]
\centering
\includegraphics[width=0.85\textwidth]{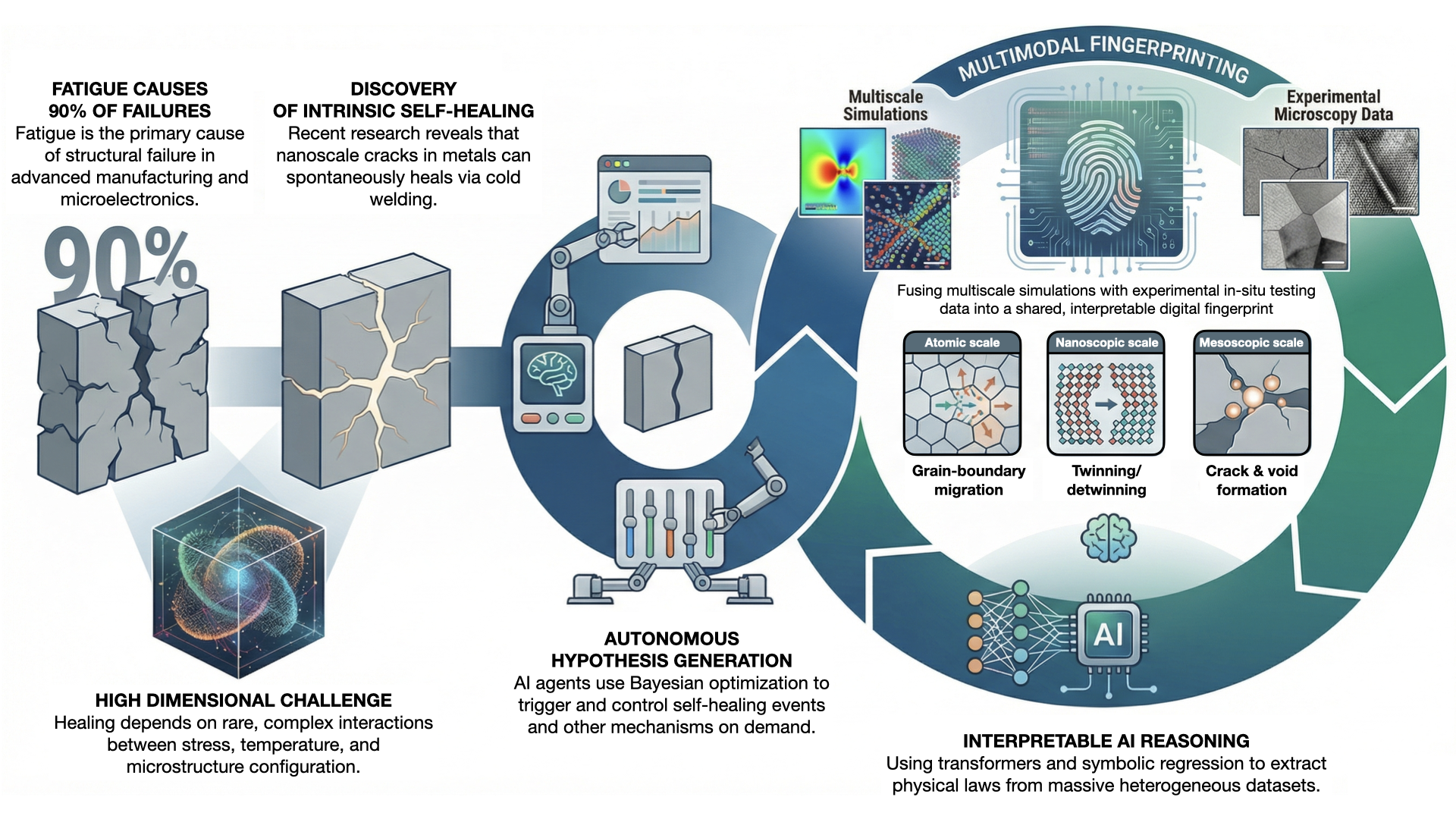}
\caption*{Graphical abstract}
\end{figure}

\noindent\textbf{Highlights}
\begin{itemize}
\item A probabilistic framework casts materials behavior as mechanism ensembles
\item Multiscale simulation, multimodal data, and ML are fused within the framework
\item Example on crack self-healing shows fatigue as inference over mechanism competition
\item The framework generalizes to emergent phenomena across physical and chemical systems
\end{itemize}

\clearpage
\input{nomenclature}

\newpage
\section{Introduction}\label{sec:intro}
Across physical and life sciences, understanding complex dynamical systems has long required statistical frameworks that capture not only the rates of individual processes, but also the competition, synergy and correlation among co-active processes.
Fokker-Planck descriptions of reaction-diffusion systems~\cite{risken1989fokker, jordan1998variational} and kinetic Monte Carlo approaches in surface chemistry~\cite{andersen2019practical} illustrate this principle: the activation of one process modifies the probability of others.
In plasticity and damage mechanics, however, this perspective has matured more slowly.
The reason is not a lack of awareness that mechanisms interact, but rather because defect interactions in crystalline solids are non-local and long-range in a way that makes the correlation problem genuinely hard: a dislocation pile-up a hundred nanometers from a grain boundary modifies its migration barrier~\cite{shen1988dislocation}; solute segregation induced by a nearby dislocation substructure both reduces the grain-boundary migration barrier~\cite{suhane2022solute} and increases its resistance to decohesion~\cite{dingreville2017primer}; opposing effects arising from the same configurational change that cannot be captured by summing individual barriers.
Such couplings mean that the effective energy landscape governing collective defect behavior cannot be reduced to a simple sum of individual barriers.

Fatigue crack propagation is one such process, emerging from the collective activation of multiple unit mechanisms operating across atomic and mesoscale length scales~\cite{pineau2016failure,mcdowell2024nonequilibrium1,mcdowell2024nonequilibrium2,mcdowell2025hierarchical}.
The relative activation rates of fatigue mechanisms depend on local microstructure and applied stimuli, and their collective dynamics determine the macroscopic response.
The central thesis of this perspective is that complex material phenomena are best understood and designed as conditional probabilistic superpositions of identifiable unit processes.
While a bottom-up enumeration of these probabilities is a Herculean task, the convergence of exascale computing capabilities, multimodal characterization, and interpretable artificial intelligence (AI) and machine learning (ML) is beginning to make inference of this conditional probability landscape tractable.

In contrast to the present era of data-centric research, the many-decades-old phenomenological approaches of materials engineering (e.g., rate-theory swelling models for irradiation~\cite{brailsford1972rate,mansur1978void}, Johnson-Mehl-Avrami-Kolmogorov (JMAK) kinetics for phase transformations~\cite{johnson1939reaction,avrami1939kinetics,kolmogorov1937statistical}, or Paris-law damage tolerance for fatigue crack propagation~\cite{suresh1998fatigue}) collapse a wealth of mechanistic information into low-dimensional, human-derived heuristics. 
While highly productive for baseline predictions, these laws remain inherently blind to regimes where the underlying mix of mechanisms shifts, or where the state of the materials system evolves far from classical thermodynamic expectations.
The closest the field has come to moving beyond this limitation is transition state theory applied to defect ensembles~\cite{mcdowell2024nonequilibrium1,mcdowell2024nonequilibrium2}, whose coarse-grained realization in models such as Kocks-Mecking law captures effective barriers for dislocation glide and reproduces macroscopic flow stress across a broad range of loading conditions.
Yet these approaches remain limited to their calibration regimes and to mechanisms treated in isolation.
What they do not represent is conditional dependencies between co-active mechanisms; instead, such dependencies are usually introduced through constitutive assumptions that prescribe interaction rules rather than infer them.
As a result, phenomenological laws such as the Paris-law constants act as effective averages over a static distribution of active mechanisms, rather than as predictors of what happens when that distribution changes.

To bridge this gap, conditional dependencies among unit mechanisms can be inferred through the fusion of accelerated multiscale simulation and multimodal experimental characterization.
In this article, we elaborate such a strategy in the context of fatigue-crack propagation in metals, though the approach presented herein is thought to be generalizable to many other domains.
The present work is therefore a forward-looking perspective: it outlines what may become possible, rather than what has already been verified.
The concepts are introduced in broad terms here, with the details to be developed in future work.

\subsection{The classical fatigue exemplar and its limits}
{
Fatigue accounts for up to 90\% of structural failures~\cite{suresh1998fatigue}.
For more than a century, damage-tolerance design has rested on a largely unchallenged assumption: crack growth is a strictly forward and irreversible process ($\mathrm{d}a/\mathrm{d}N > 0$, where $a$ is the crack length and $N$ the number of cycles).
Engineering analysis is therefore directed at characterizing the rate of damage accumulation through empirical laws such as the Paris law ($\mathrm{d}a/\mathrm{d}N$ vs.\ $\Delta K$, where $K$ is the stress intensity factor and $\Delta K$ its cyclic range), implemented in deterministic (NASGRO~\cite{mettu1999nasgro}) and probabilistic (DARWIN~\cite{mcclung2018probabilistic}) codes.
These tools have proven to be predictive, but because they operate at the scale of macroscopic crack advance, they provide no direct access to the atomistic and mesoscale processes that determine whether a crack grows, stalls or deviates.
The Paris-law constants therefore function as effective averages over a static distribution of active mechanisms, collapsing mechanistic diversity into a single empirical relation at the expense of physical transparency and interpretability.
As a result, scatter in legacy measurements is absorbed as measurement variability rather than interpreted as information, obscuring the tails of the distribution in which exceptional fatigue-resistance may reside.
}

\subsection{Self-healing as a proof of concept}
\begin{figure}[!htbp]
  \centering
  \includegraphics[width=0.99\textwidth]{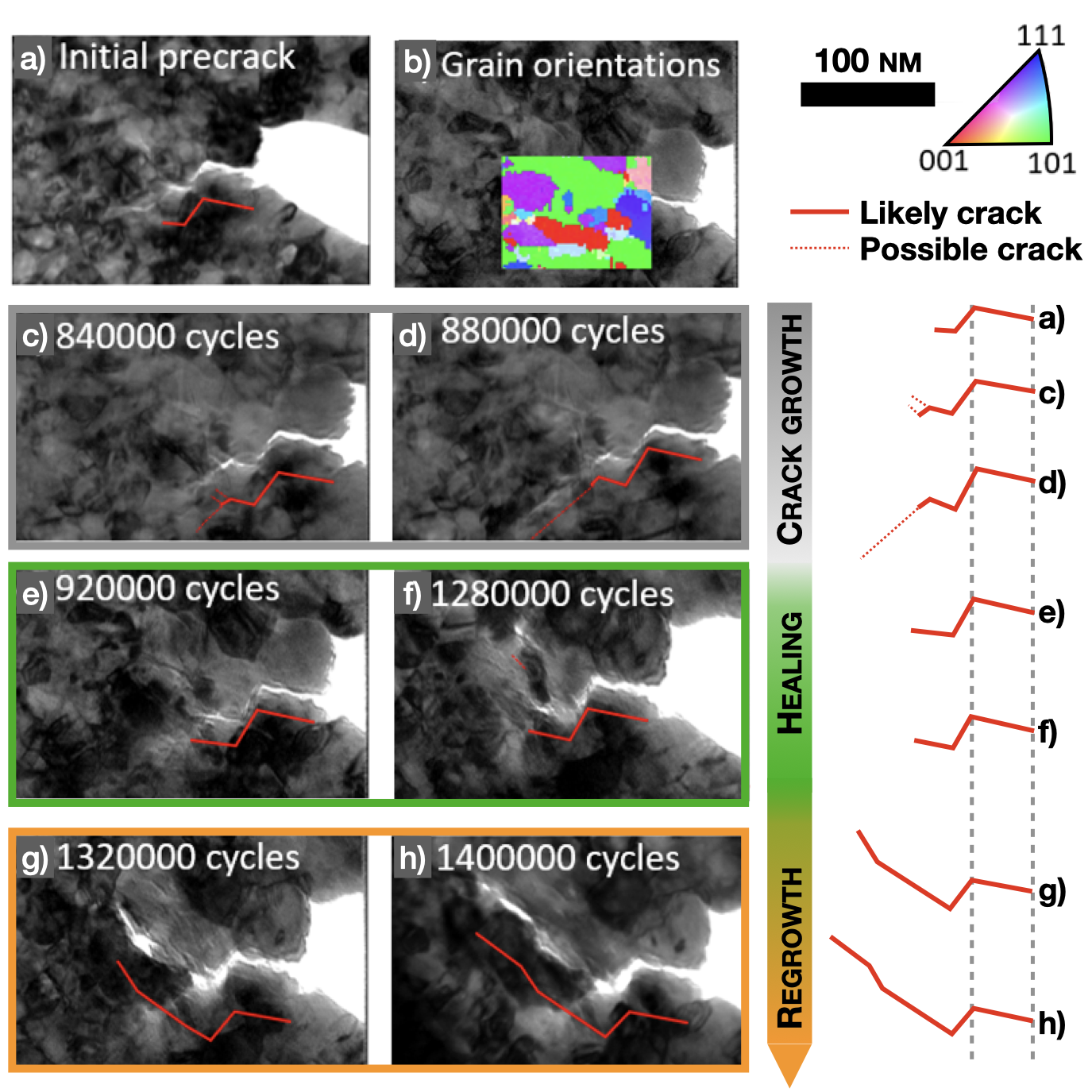}
  \caption{{\bf Crack self healing in nanocrystalline metals.} Progression of in-situ transmission electron microscopy (TEM) still images showing nanoscale fatigue
crack growth and healing as a function of the number of cylces in Pt~\cite{barr2023autonomous}, driven by multiple parallel and sequential processes. (a) Initial precrack configuration; (b) grain orientation distribution; (c)-(d) early crack growth events; (e)-(f) crack healing events; (g)-(f) crack regrowth events.}
  \label{fig:selfhealing}
\end{figure}

{
Barr et al.~\cite{barr2023autonomous} exposed a fundamental blind spot in this irreversible picture (see Fig\@.~\ref{fig:selfhealing}).
Using a MEMS-based, in-situ transmission electron microscopy (TEM) fatigue platform~\cite{bufford2016high,pierron2006methodology}, they directly observed crack self-healing in nanostructured platinum under high-cycle cyclic loading: propagating cracks arrested and partially healed before re-initiating along new paths.
This emergent phenomenon was attributed to compressive strains from stress-driven grain-boundary migration, which brought the crack flanks into cold-welding contact. 
Comparable emergent self-healing events were also documented in nanostructured copper~\cite{barr2023autonomous}.
These observations do not invalidate damage-tolerance for engineering alloys, but they do show that the assumption $\Delta a > 0$ does not always hold.
More broadly, the experimental observations provide direct evidence that, under specific microstructural and loading conditions, mechanism competition can favor crack reversal over continued propagation.
}

\begin{figure}[!htbp]
  \centering
  \includegraphics[width=0.99\textwidth]{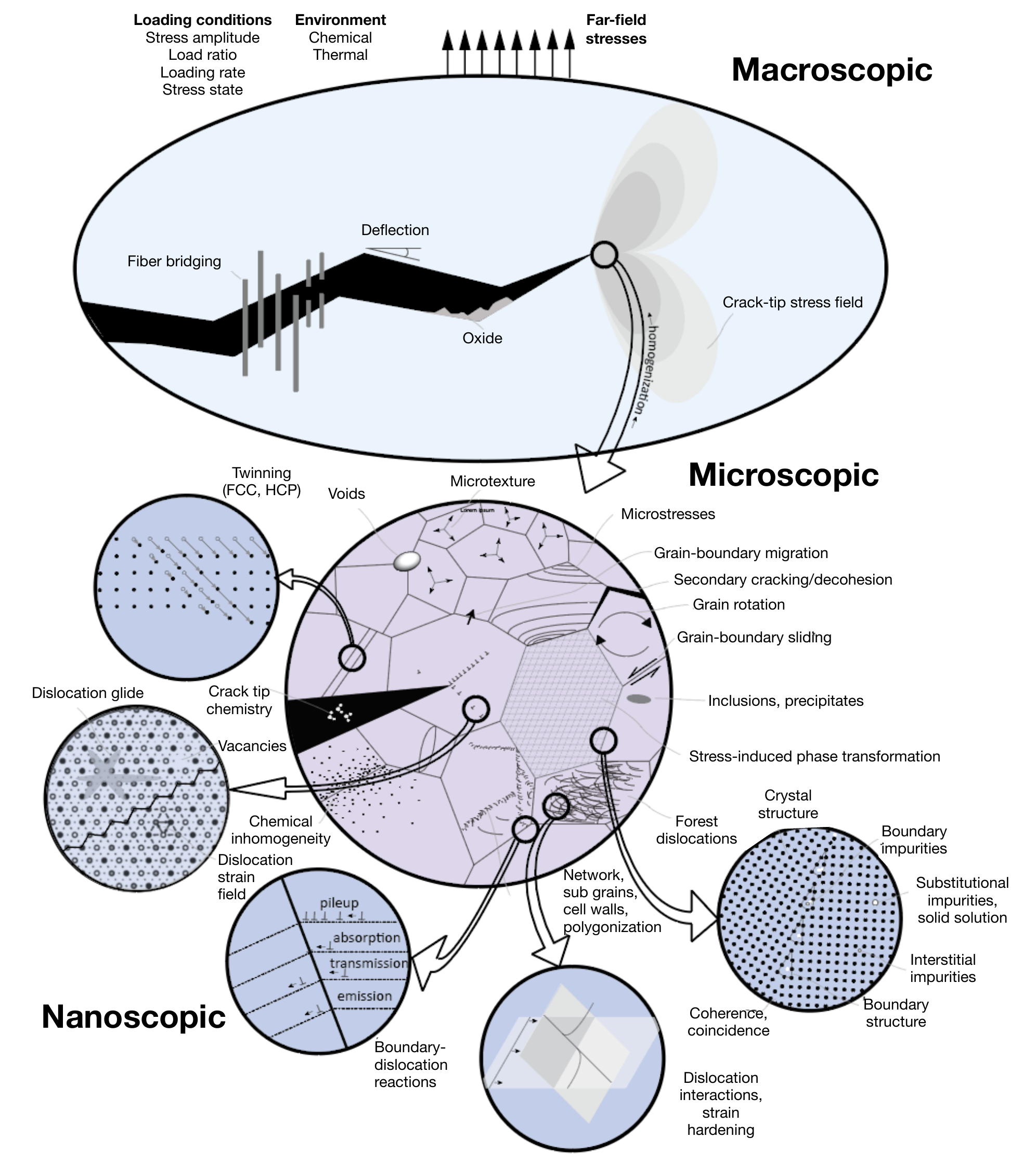}
  \caption{{\bf Hierarchical decomposition of the constituent processes and structural features influencing fatigue crack propagation across macroscopic, microscopic, and nanoscopic length scales.} Each process at a given scale emerges from an ensemble of finer-scale constituents, illustrating the scale-relative character of the mechanistic vocabulary developed in this section.}
  \label{fig:gazinta}
\end{figure}

{
Nanoscale self-healing highlights a class of behavior that current theories are not designed to represent.
It reflects local interactions among multiple unmodeled mechanisms and implies that fatigue cracking is better viewed as an emergent dynamical process than as a single monotonic mechanism.
In this picture, crack advance, arrest and healing arise from specific combinations and sequences of constituent mechanisms operating at the grain, grain-boundary and dislocation scales (Figure~\ref{fig:gazinta}).
}

{
From this perspective, cumulative fatigue-damage indicators should be interpreted as collective signatures of multi-mechanism activity across scales.
The key challenge is to connect what can be observed experimentally (i.e., crack healing under tensile load) with the underlying mechanism interactions that produce it (i.e., the collective interaction between mechanisms leading to resorption of the crack).
Addressing that gap requires quantitative learning of a conditional probability landscape for mechanism competition.
Recent advances in exascale computing, accelerated materials simulations with machine learning~\cite{montes2022training,galvelis2023nnp}, and foundation models for multimodal data fusion~\cite{mccabe2023multiple} have begun to remove three longstanding barriers:
the timescale gap between detailed atomistic and mesoscale simulation and experimental fatigue,
the configuration-space dimensionality that prevented comprehensive sampling of microstructural conditioning variables, and
the absence of a shared representation capable of fusing mechanistically labeled simulation data with partial experimental observables.
We therefore recast the classical materials science `process-structure-property-performance' (PSPP) paradigm as an actionable Discover-Learn-Interpret-Elicit (DLIE) scientific cycle~\cite{tsao2024ai}, better suited to integrating AI/ML tools with materials-science instruments and techniques.
The components of this cycle are introduced below, and the probabilistic framework is later extended to other physical and chemical domains in which emergent processes are prevalent.
}

\section{Constituent Mechanisms vs\@.~Observables}
\label{sec:const-mech-obs}
A central challenge in multiscale materials science is to relate the discrete physical mechanisms that drive the mechanical response to the partial macroscopic observables available to characterize them.
Because constituent unit mechanisms are often conceptualized differently across modeling and experimental communities and across spatiotemporal scales~\cite{dingreville2016review}, a consistent, scale-relative nomenclature that
unifies physical processes with their measurable signatures is a prerequisite
for the probabilistic framework developed in subsequent sections.
All the symbols and terms used hereafter are defined in the nomenclature table provided at the beginning of this article.

\subsection{A scale-relative mechanistic vocabulary}
{
In fatigue, conventional alloys often conceal localized unit processes beneath the empirical averages of the Paris law, whereas the cyclic self-healing response of nanostructured metals departs markedly from this baseline and provides a sensitive regime for developing a probabilistic framework~\cite{polak2017profiles,chowdhury2016mechanisms}.
When the grain size decreases into the range of tens to hundreds of nanometers, classical dislocation-dominated plasticity gives way to an ensemble of boundary-mediated deformation and damage pathways~\cite{mughrabi2010cyclic,padilla2010review}:
dislocation mean free paths shrink below the characteristic grain diameter, favoring stable intragranular substructures while at the same time suppressing persistent slip band development until fatigue-induced grain coarsening locally reactivates standard slip mechanisms~\cite{furnish2018evidence,chen2022heterogeneous}, while the grain boundary transitions from a passive kinematic obstacle to an active fatigue participant capable of sliding, migrating, emitting dislocations, or undergoing localized decohesion.
}

{
We use the term \emph{constituent mechanism} in a strictly scale-relative sense: mesoscale features are treated as compound processes decomposable into finer-scale events at lower rungs of the multiscale hierarchy, eventually resolving into atomic-scale fundamental transitions (Figure~\ref{fig:gazinta}).
The term ``constituent'' therefore denotes operational identifiability and physical distinctiveness at a chosen level of abstraction, rather than absolute thermodynamic irreducibility.
We denote individual mechanisms by \(\mathcal{O}_i\), where the index $i$ labels a specific process at the chosen level of the hierarchy.
The same symbol is used at finer scales when a given \(\mathcal{O}_i\) is itself decomposed into sub-mechanisms, reflecting the recursive character of the vocabulary.
Prior experiments and simulations have detailed the specific constituent mechanisms governing cyclic loading in nanostructured metals~\cite{mughrabi2010cyclic,qiu2024grain,furnish2018evidence,lu2021fatigue,farkas2005atomistic,chen2022heterogeneous,padilla2010review,jain2026putting}.
Examples include
stress- and curvature-driven grain-boundary and twin-boundary migration~\cite{qiu2024grain};
fatigue-induced abnormal grain growth that concentrates strain at triple junctions~\cite{furnish2018evidence,lu2021fatigue};
nanovoid nucleation and coalescence via vacancy aggregation along interfaces~\cite{farkas2005atomistic};
grain-boundary decohesion under local tensile traction~\cite{pineau2016failure_b};
interfacial sliding accommodated by atomic shuffling and dislocation emission from boundary ledges~\cite{padilla2010review}.
Each constituent path operates as a thermally activated, stochastic process whose kinetics are dictated by the local atomic arrangement, boundary chemistry, and the instantaneous stress--temperature state.
}

\subsection{Conditioning variables and partial observables}
{
A defining feature of this constituent-mechanism ensemble is that these processes are neither mutually exclusive nor strictly sequential: they may occur simultaneously or in sequence, and in synergistic, antagonistic or independent combinations.
The probability of activating a given mechanism \(\mathcal{O}_i\) is conditioned on the local subsystem state \(\mathcal{S}\), which encodes the local microstructural configuration, defect population and loading history, and on the global conditioning variables \(\mathcal{M}\), which captures the microstructure (e\@.g\@.~crystallographic texture) and long-range elastic and plastic fields that shape the local energy landscape.
Grain size is a primary conditioning variable: as the mean grain diameter decreases from $\sim$100~nm toward $\sim$10~nm, grain-boundary migration and sliding become more probable, while dislocation-mediated plasticity and persistent slip band formation become less likely, consistent with the geometric suppression of intragranular slip and the increased relative area of grain boundaries.~\cite{mughrabi2010cyclic,padilla2010review}.
Grain-boundary character is a second critical variable: boundaries near special coincidence site lattice (CSL) orientations, particularly $\Sigma3$ coherent twin boundaries, have lower migration barriers and higher resistance to decohesion than general high-angle boundaries, so the microstructural distribution of boundary types directly modulates which mechanisms are thermodynamically accessible at a given stress amplitude~\cite{qiu2024grain}.
Pre-existing defect states, such as
the dislocation density inherited from processing,
the population of vacancies and nanovoids accumulated in prior loading, and
the degree of solute segregation at boundaries,
further shift the conditional activation probabilities by modifying local free-energy barriers.
Among applied loading variables, stress amplitude controls the driving force for all stress-activated mechanisms;
load ratio $R = \sigma_{\min}/\sigma_{\max}$ governs the degree of stress reversal and thereby the likelihood of mechanisms such as twinning/detwinning or crack-flank contact that depend on compressive excursions;
loading frequency modulates the competition between athermal and thermally activated diffusive processes; and
temperature shifts the entire activation energy landscape through Boltzmann weighting.
Because the operative mechanism mix is determined by the full joint distribution of these microstructural and loading variables rather than any single one, their systematic sampling motivates accelerated multiscale simulation strategies
and multimodal experimental campaigns (Sections~\ref{sec:acl-multiscale-sim} and~\ref{sec:exp-sim-sig}).
} 
 
{
The experimental observables available to characterize this ensemble do not report on individual mechanisms directly, but rather on their collective, integrated effects.
For instance, surface topography measured by atomic force microscopy (AFM) or scanning electron microscopy (SEM) encodes the morphological imprint of cumulative plastic irreversibility through extrusions, intrusions, and ridge-and-valley patterns associated with persistent slip bands and grain-boundary migration~\cite{polak2017profiles,payam2019development}. 
Digital image correlation (DIC) resolves the localized strain discontinuities at individual grain boundaries that are predictive of fatigue crack initiation~\cite{stinville2022origins}.
Electron backscatter diffraction (EBSD) reveals grain coarsening patterns and crystallographic texture changes associated with
stress-driven boundary migration~\cite{furnish2018evidence}.
In-situ TEM using nanomechanical loading platforms~\cite{bufford2016high,pierron2006methodology} provides direct visualization of individual mechanism events such as grain-boundary migration steps, dislocation emission bursts, and nanovoid nucleation, but at the cost of severe geometric and throughput constraints that limit statistical inference.
Faster, indirect surrogate probes are also possible; second-harmonic generation (SHG) optical microscopy, for instance, offers a potentially high-throughput alternative to classical microstructural characterization; its sensitivity to symmetry-broken interfacial environments suggests possible mechanistic specificity for grain boundaries and dislocation substructures~\cite{yokota2012optical,shafiei2021detection,hristu2014nonlinear,bozhevolnyi2003direct,prylepa2018material,rellaford2021characterization}, though its utility for fatigue damage monitoring in nanocrystalline metals remains to be demonstrated.
Each modality therefore provides a partial view of the mechanism space: insufficient on its own to reconstruct the full multi-mechanism picture, but, when fused with mechanistic simulation data in a shared representation, sufficiently informative to support quantitative inference (Section~\ref{sec:exp-sim-sig}).
}

{
The emergent fatigue processes of engineering interest, including crack nucleation, propagation, arrest and self-healing, each arise from specific sequences and combinations of these constituent mechanisms conditioned on local microstructural state~\cite{furnish2018evidence,farkas2005atomistic,pineau2016failure,barr2023autonomous}.
Simulation modalities~\cite{qiu2024grain,bamney2023assessing2,farkas2005atomistic,xu2013healing,chakraborty2024role,bieberdorf2023grain2,miehe2010thermodynamically,miehe2010phase,duda2015phase,miehe2016phase,svolos2022fourth,svolos2025phase,livingston2025inference} can isolate individual mechanisms with high fidelity but presuppose rather than predict their activation rates.
Experiments project the full hierarchy onto partial observables.
Neither alone reconstructs the complete joint probability landscape.
The constituent-mechanism vocabulary introduced here provides the physical basis for the probabilistic framework developed below.
The probabilistic formalism in the next section supplies its mathematical structure (Section~\ref{sec:prob-framework-mech}).
}

\section{A Probabilistic Framework for Mechanism Interaction}
\label{sec:prob-framework-mech}

{
The preceding section established that emergent fatigue behaviors arise from the collective, conditional activation of competing unit processes rather than from any single dominant mechanism.
The central question is then how to represent that collective reality quantitatively, not by prescribing how mechanisms interact, which is what every constitutive model does, but by treating the topology of those interactions as a scientific unknown to be inferred. 
Conditional probability, probabilistic graphical models~\cite{koller2009probabilistic,kalidindi2019bayesian,walker2024flow, robertson2026probabilistic} and transition state theory~\cite{mcdowell2024nonequilibrium1, mcdowell2024nonequilibrium2, mcdowell2025hierarchical} provide the natural language for this:
they encode stochastic mechanism activation,
capture conditional dependencies among co-active processes without assuming independence, and
support systematic coarse-graining across the scales at which different mechanisms operate.
This section introduces such a probabilistic framework.
Fatigue serves as the running exemplar, but the logical structure of this probabilistic framework is material- and phenomenon-agnostic.
The same logic underlies the use of master equations in chemical kinetics, Fokker-Planck descriptions of Brownian motion and reaction-diffusion systems, and kinetic Monte Carlo in heterogeneous catalysis and defect evolution: whenever a system's trajectory is determined by competing stochastic transitions among identifiable states, the right language is probability.
}

{
The central departure from existing constitutive approaches is epistemic: the topology of conditional dependencies among mechanisms (i.e. which $\mathcal{O}_i$ are correlated, which are independent given $\mathcal{S}$, and how these dependencies shift with loading and the global environment) is treated here as a scientific unknown to be learned from data, not as a modeling assumption to be prescribed in advance.
Every phenomenological law encodes coupling assumption \textit{a priori}: the Kocks-Mecking law prescribes how forest hardening couples to flow stress, the Paris law prescribes how crack-tip plasticity averages over the mechanism distribution, leaving no room to discover couplings that physical intuition did not anticipate.
The probabilistic framework described below treats those couplings as latent variables inferred through Bayesian methods, making the conditional dependency structure itself a scientific output rather than an input.  
}

\subsection{Conditional probability perspective and its  grounding in transition state theory}

We frame this perspective around three considerations.
First, the stochastic nature of materials makes uncertainty quantification indispensable.
Second, materials behavior is multiscale by construction, with important processes spanning orders of magnitude in length and time. 
We suggest, however, that the defining challenge in fatigue is the density of this scale space: many relevant mechanisms are not separated cleanly by scale, but instead coexist at similar, though not identical, length and time scales.
This motivates placing scale transitions, including coarse graining across orders of magnitude and scale concurrence, namely the coexistence of multiple mechanisms within a single scale, at the center of the discussion.

\begin{figure}[htbp]
  \centering
  \includegraphics[width=0.99\textwidth]{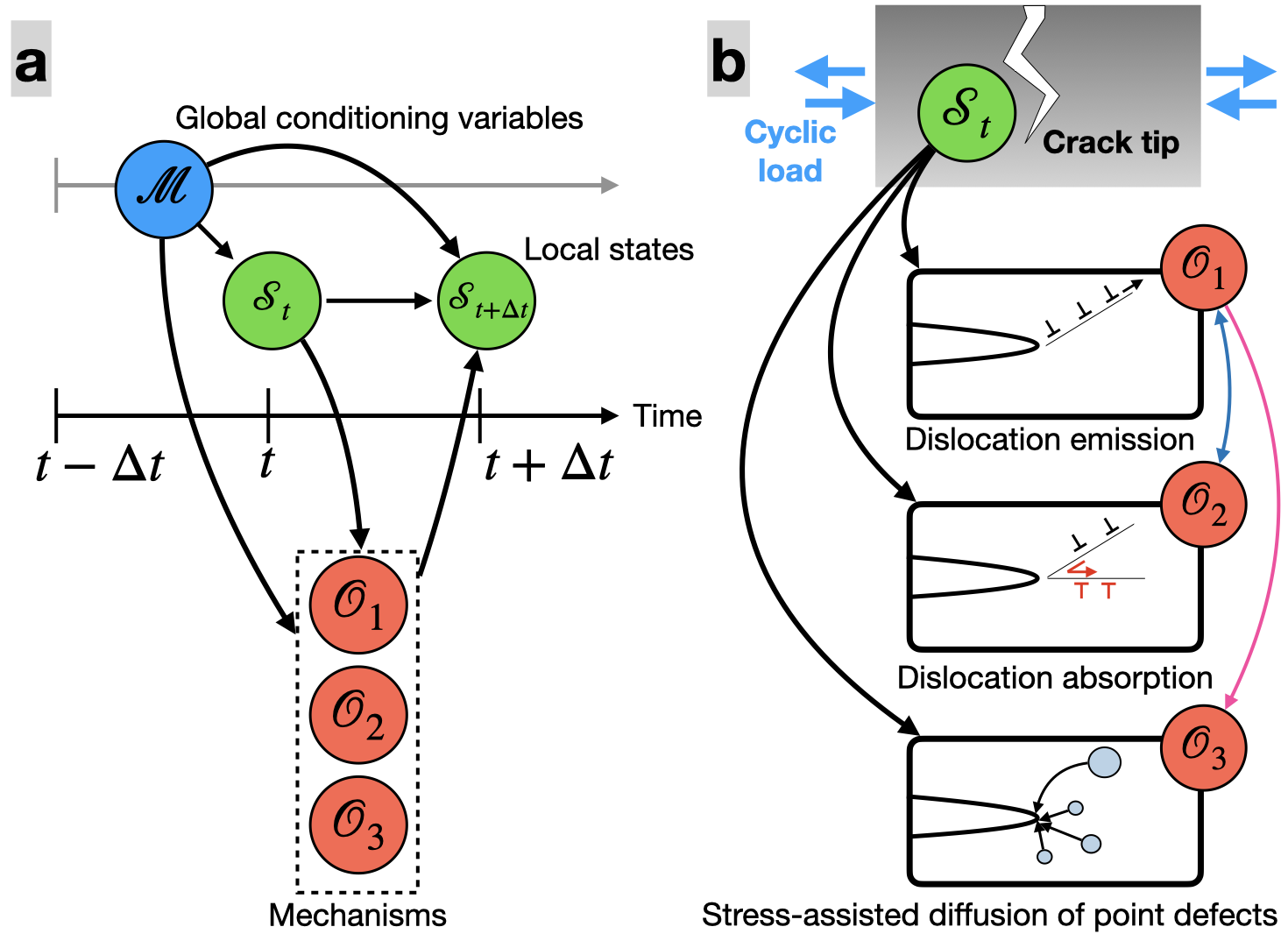}
  \caption{{\bf Probabilistic framework for mechanism interaction.}
  (a) Directed acyclic graph model representation of the conditional probability structure. Nodes represent the global conditioning variables $\mathcal{M}$, the local subsystem state at successive time steps $\mathcal{S}$, and the co-active constituent mechanisms $\{\mathcal{O}_i\}$. Arrows denote conditional dependencies of the component(s) at the tail to the one(s) at the head; the dashed box represents a series of fully interconnected mechanisms.
  (b) Physical instantiation for the minimal crack-tip system of Section~\ref{sec:ped-ex}. Three mechanisms are shown:
  $\mathcal{O}_1$ forward dislocation emission on the primary slip system;
  $\mathcal{O}_2$ dislocation absorption on a non-coplanar secondary system;
  $\mathcal{O}_3$ stress-assisted diffusion of point defects.
  Coupling arrows in (a) indicate state-mediated interactions: activation of $\mathcal{O}_1$ modifies local back-stress and thereby the accessibility of $\mathcal{O}_3$, so that $P(\mathcal{O}_3 | \mathcal{S}_{t+\Delta t}, \mathcal{M}) \neq P(\mathcal{O}_3 | \mathcal{S}_{t}, \mathcal{M})$ These couplings are inferred rather than prescribed.}
  \label{fig:DAG}
\end{figure}

Given the definitions of mechanisms $\mathcal{O}_i$, local subsystem state $\mathcal{S}$, and global conditioning variables $\mathcal{M}$ in Section~\ref{sec:const-mech-obs} and in the Nomenclature table, the probabilistic framework is built from three coupled conditional distributions that link mechanism activation, state evolution and macroscopic observables across scales (Fig.~\ref{fig:DAG}). 
$P(\{\mathcal{O}_i\}_{i=1}^N \mid \mathcal{S}_t, \mathcal{M})$ defines the joint probability that a finite set of identified mechanisms co-occurs conditioned upon the local state and global variables, that is, which unit processes are thermodynamically and kinetically accessible under the current microstructural and loading state, and in what combinations. $P(\mathcal{S}_{t+\Delta t} \mid \{ \mathcal{O}_i \}_{i=1}^N, \mathcal{S}_t, \mathcal{M})$ describes how the state transitions between $t$ and $t + \Delta t$ also conditioned upon activation of the identified mechanisms and global variables, capturing the kinematic and thermodynamic consequences of an event sequence, including changes in dislocation density, boundary configuration and crack length, and thereby the mechanisms that become accessible at the next step.
Finally, $P(E | \mathcal{S}_{0:\tilde{t}}, \mathcal{M}_{0:\tilde{t}})$ is the probability of a macroscopic observable conditioned upon time sequences of states and global variables, providing the bridge between mechanism-level descriptions and experimental measurements, which typically integrate over many mechanism activations rather than resolve any single event.
Here, $\mathcal{S}_{0:\tilde{t}}$ refers to the sequence of states over instants $\{0,...\tilde{t}\}$.
A central goal is to identify independence relations within these distributions that enable simplification, and to develop efficient strategies for calibrating the resulting approximations from the heterogeneous simulation and experimental data described in Sections~\ref{sec:acl-multiscale-sim} and~\ref{sec:exp-sim-sig}.
This formulation rests on several working assumptions: 
that $\mathcal{M}$ evolves more slowly than $\mathcal{S}$,
that mechanisms can be identified and labeled from simulation or experiment, and 
that the chosen time discretization is long compared with individual events but short compared with macroscopic damage accumulation.
Each of these assumptions remains an open question in the community.

The framework above captures two coupled operations:
it models evolution at a chosen scale while also accounting for information transfer across scales.
Information from higher scales enters through the global conditioning variables $\mathcal{M}$), whereas information from lower scales enters through the definition of the mechanisms $\mathcal{O}_i$ and the approximation of the associated probability distributions (Fig.~\ref{fig:DAG}a).
We therefore view $\mathcal{O}_i = c_i(s_{t:t+\Delta t}, \mathcal{S}_t, \mathcal{M})$ as a coarse-graining operator that extracts the occurrence of mechanism $\mathcal{O}_i$ from lower-length-scale states ${s}_{t:t+\Delta}$ over the time interval $\left[t, t+\Delta t \right]$, conditioned on $\mathcal{S}_t$ and $\mathcal{M}$.
The principal challenge lies in identification, namely, disentangling complex data streams to isolate unit mechanistic transitions in lower-scale experiments and simulations.
Once identified, these events must still be approximated statistically, a task that requires advances in active exploration~\cite{chen2017beyond, montes2022training} and data generation~\cite{robertson2024micro2d, buzzy2025polymicros}.

\subsection{A Pedagogical Example}\label{sec:ped-ex}

To illustrate the probabilistic framework described above, we consider a minimal crack-tip system $\Sigma$ extracted from a polycrystalline metal.
The global conditioning variables $\mathcal{M}$ represent the applied cyclic traction, the thermal-chemical environment and the microstructure (grain size, texture, etc.), which together define the far-field boundary conditions acting on $\Sigma$.
The local subsystem state $\mathcal{S}$ is described by a coarse set of variables that includes the mobile and forest dislocation densities ($\rho_m, \rho_f$), the neighboring void density ($d_v$), and the crack length ($a$), $\mathcal{S} = \{\rho_m, \rho_f, d_v, a \}$.
For the sake of illustration, we restrict our attention to three constituent mechanisms $\{\mathcal{O}_i\}_{i=1}^3$:
$\mathcal{O}_1$, dislocation emission at the crack tip by forward slip on the primary system;
$\mathcal{O}_2$, dislocation absorption at the crack tip by secondary slip on a non-coplanar system; and
$\mathcal{O}_3$, point defect stress-assisted diffusion (Fig\@.~\ref{fig:DAG}b).
This set is not exhaustive, but it provides a minimal closure that includes the mechanisms implicated in nanostructured-metal self-healing~\cite{barr2023autonomous} as well as those commonly associated with conventional fatigue damage~\cite{mughrabi2010cyclic, padilla2010review, pineau2016failure}.

In this pedagogical example the mechanisms are defined explicitly to illustrate the construction of the probabilistic framework, but in realistic applications they would instead be identified by applying the coarse-graining operators $c_i$ introduced above to simulation trajectories or experimental data streams.

The central point is that the mechanisms do not act independently and illustrates why a probability framework is necessary: activation of one mechanism alters $\mathcal{S}_t$, which in turn changes the accessibility of the others.
For example, activation of $\mathcal{O}_1$ (dislocation emission) increases the local dislocation content and back-stress, which can simultaneously raise the barrier for further forward slip and change the driving force for stress-assisted diffusion of point defects.
In other words, $P(\mathcal{O}_3|\mathcal{S}_{t+\Delta t}, \mathcal{M})$ differs from $P(\mathcal{O}_3|\mathcal{S}_{t}, \mathcal{M})$.
It is precisely this state-mediated coupling among mechanisms that a conditional-probability formalism is designed to represent and infer.

Time is discretized on a interval $\Delta t$ chosen to be long compared with individual mechanism events but short compared with macroscopic damage accumulation.
In cyclic fatigue, $\Delta t$ may correspond to one cycle or a small number of cycles.
Under this discretization, the probabilistic model is populated by estimating the joint activation distribution $P(\{\mathcal{O}_i\}|\mathcal{S}_{t}, \mathcal{M})$ and the state-transition distribution $P(\mathcal{S}_{t+\Delta t}|\{\mathcal{O}_i\}, \mathcal{S}_{t}, \mathcal{M})$.
These conditional distributions are to be estimated by multiscale simulation, theoretical rate models, and experimental observation.

A convenient starting point for mechanistic parametrization is a thermally activated rate model.
For each mechanism $\mathcal{O}_i$, one may write a free‑energy barrier $\Delta G_i(\mathcal{S}_t, \mathcal{M}) = \Delta H_i(\mathcal{S}_t, \mathcal{M}) - T\,\Delta \textbf{s}(\mathcal{S}_t, \mathcal{M})$.
Here, the enthalpy term($\Delta H_i$) describes the effect of work done (e.g., stress $\sigma$) and the entropy ($\Delta \textbf{s}$) collects the configurational and vibrational entropy of the lower length scale objects involved in the mechanism.
With the Gibbs free-energy, we can define the thermally activated rate:
\begin{equation}
    \dot{p}_{\mathcal{O}_i} = \nu_{0,i}\,N_i\,\exp\!\left[-\frac{\Delta G_i(\mathcal{S}_t, \mathcal{M})}{k_B T}\right],
    \label{eq:rate_i}
\end{equation}
where $\nu_{0,i}$ is an attempt frequency, $N_i$ enumerates equivalent sites for the event, and the selected time discretization $\Delta t$ determines the expected event count $\lambda_i = \dot{p}_{\mathcal{O}_i}\Delta t$.
As a first approximation, and when dependencies are weak over $\Delta t$, the event counts may be modeled as a Poisson distribution with expectation $\lambda_i$.
This representation is useful for rapid forward sampling and for designing targeted simulation campaigns to populate the mechanism‑space statistics.

When defect species interact strongly, the effective energy landscape is not reducible to a sum of single‑mechanism barriers.
In these regimes mesoscale ``super‑energetic'' basins emerge~\cite{mcdowell2024nonequilibrium1,mcdowell2024nonequilibrium2}: hierarchical energy structures whose topology encodes
which combinations of constituent mechanisms are collectively stable,
which are transient, and
which generate macroscopically observable outcomes such as crack arrest or self‑healing.
Constructing these basins requires upscaling atomic‑scale rates while preserving conditional dependencies among co‑active mechanisms; this upscaling is a principal objective of the accelerated multiscale simulation strategies described in Section~\ref{sec:acl-multiscale-sim}.

The practical value of the model is that macroscopic observables are obtained by marginalizing the coupled conditional distributions over mechanism activations, local states, and their transitions.
For a given externally imposed $\mathcal{M}$, the expected crack-length increment per cycle is:
\begin{equation}
\left\langle \frac{\Delta a}{\Delta N} \right\rangle_{\!\mathcal{M}}
=
\iiint
\bigl[a(\mathcal{S}_{t+\Delta t}) - a(\mathcal{S}_{t})\bigr]\,
P\!\left(\mathcal{S}_{t+\Delta t} \mid \{\mathcal{O}_i\}, \mathcal{S}_t, \mathcal{M}\right)
P\!\left(\{\mathcal{O}_i\} \mid \mathcal{S}_t, \mathcal{M}\right)
P\!\left(\mathcal{S}_t \mid \mathcal{M}\right)
\,d\{\mathcal{O}_i\}\,d\mathcal{S}_t\,d\mathcal{S}_{t+\Delta t}.
\end{equation}

A fully unconditional expectation, obtained by further marginalizing over $P(\mathcal{M})$,
is in principle possible but generally intractable -- analogous to the marginal likelihood
in Bayesian inference -- and is not required when $\mathcal{M}$ is prescribed by the
experimental or operational conditions.
In the single-cycle discretization limit, this relation recovers a probabilistic generalization of the Paris law, in which the classical crack-growth rate emerges as the expectation of $\Delta a$ marginalized over all mechanism combinations and microstructural states consistent with the applied loading.
The Paris-law constants $C$ and $m$ (in $\mathrm{d}a/\mathrm{d}N = C\,\Delta K^{m}$) therefore appear as fits to the expectation over the mechanism and microstructural distributions sampled during calibration, rather than as fundamental, scale-independent material parameters.
More importantly, the variance of $\Delta a$ across realizations, which the classical Paris law suppresses entirely, encodes the mechanistic heterogeneity introduced by the probabilistic process and responsible for tail phenomena, including crack arrest and self-healing.
These moments, together with the independence structure that supports tractable approximations, are key to a predictive probabilistic description of fatigue.

To summarize, the pedagogical example above illustrates three practical points.
First, the operators $c_i$ define mechanism labels and compress lower-scale dynamics into the statistics consumed by the probabilistic model.
Second, mechanism activations update the local state $\mathcal{S}$ and thereby mediate conditional dependencies among mechanisms; these couplings must be inferred rather than prescribed.
Third, once populated, the coupled conditional distributions permit forward simulation and inverse inference that connect mechanism-level dynamics to macroscopic observables of engineering relevance.
The subsequent sections detail the computational, experimental, and inferential
machinery needed to realize these objectives and quantify the uncertainty accompanying
each stage of the pipeline.

\section{Accelerated Multiscale Simulation: Populating the Mechanism Space}
\label{sec:acl-multiscale-sim}

Once the conditional probabilistic structure is established, the next challenge is to populate it with mechanistically labeled data, including systematic samples of energy barriers, kinetic rates, and branching ratios across relevant microstructural configurations, loading conditions, and length scales.
These data can then be assembled into the conditional distributions governing mechanism activation, state evolution, and emergent outcome.
Accelerated simulations including accelerated molecular dynamics (AMD) methods~\cite{plimpton2020parallel,zotov2022entropy,kastner2011umbrella,zamora2016modern,voter1998accelerating,shaw2009millisecond}, machine-learning interatomic potentials~\cite{zuo2020performance,musil2021physics,goff2024permutation,choudhary2024jarvis,rohskopf2023fitsnap,wood2019data,bartok2018machine,daw2023simple,mendelev2008analysis}, discrete dislocation dynamics~\cite{sobie2015analysis,sobie2017scale,sobie2017thermal,capolungo2019gd3} and phase-field methods~\cite{montesdeoca2021accelerating,oommen2022learning,oommen2024rethinking,dingreville2024benchmarking}, provides the means to do so by identifying candidate mechanisms, estimating their barriers and rates, and translating those local statistics into a computable probability landscape.
In this setting, atomistic and mesoscale methods are not alternative descriptions of the same phenomenon, but complementary components of a hierarchical inference pipeline: the former resolve rare events and barrier-crossing pathways, whereas the latter extend those statistics to collective defect evolution, crack advance and boundary-mediated relaxation processes.
Although molecular dynamics can resolve individual unit mechanisms, it cannot by itself span the full design space of nanocrystalline fatigue, which includes grain sizes of roughly 10 to 100 nm, diverse grain-boundary character, and broad ranges of stress amplitude and temperature.
Upscaling over space and time and concurrent modeling are therefore essential for transferring information across scales and building a computable probabilistic description that can support scale-aware prediction, active exploration of under-sampled states and, ultimately, a probabilistic description of emergent fatigue behavior.

\begin{figure}[htbp]
  \centering
  \includegraphics[width=0.99\textwidth]{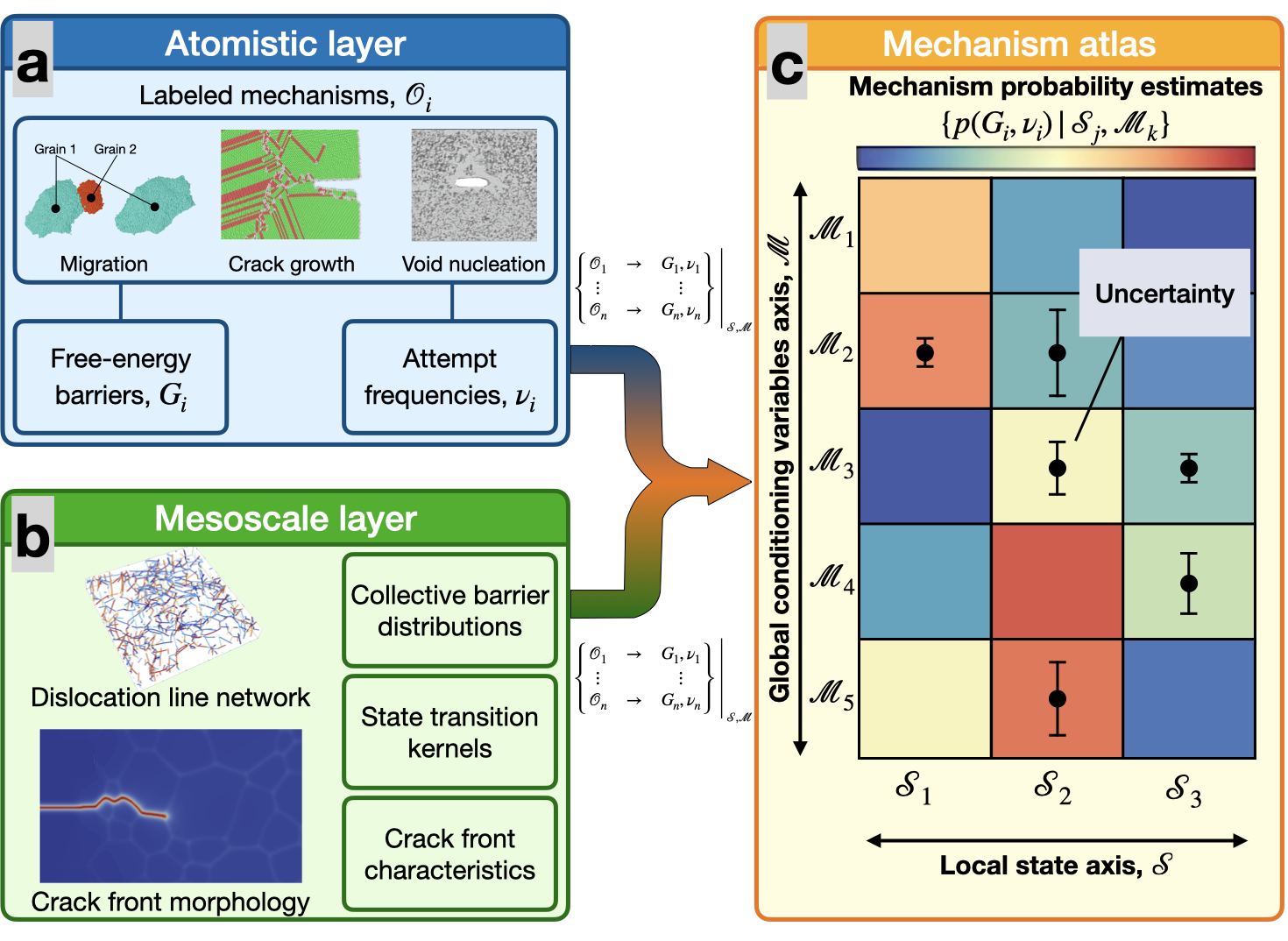}
  \caption{{\bf Hierarchical simulation pipeline for populating the conditional probability landscape of mechanism activation across atomistic and mesoscale scales.}
  (a) Atomistic layer: accelerated molecular dynamics (AMD) and machine-learning interatomic potentials (MLIAPs) extract free energy barriers $G_i$, and attempt frequencies $\nu_i$ across systematically varied grain-boundary types, grain sizes, crack configurations, stress amplitudes, and temperatures, yielding conditional mechanism probabilities $P(\mathcal{O}_i | \mathcal{S},\mathcal{M})$. 
  (b) Mesoscale layer: phase-field and discrete dislocation dynamics (DDD) simulations translate local atomistic activation events into collective barrier distributions and state-transition statistics.
  (c) Mechanism atlas: the integrated output of the atomistic and mesoscale layers, structured as an uncertainty-quantified dataset of labeled mechanism activation events, effective barrier distributions, branching ratios, and state-transition statistics spanning the relevant design space of local states and global conditions.}
  \label{fig:multiscale}
\end{figure}

\subsection{Atomistic layer: energy barriers and mechanism kinetics}\label{sec:atomiclayer}
At the atomistic scale, the probabilistic framework in Section~\ref{sec:prob-framework-mech} requires the Gibbs free-energy barrier Gibbs $\Delta G_i(\mathcal{S},\mathcal{M})$, attempt frequencies $\nu_{0,i}$, and the multiplicity $N_i(\mathcal{S})$ associated with each mechanism $\mathcal{O}_i$.
These quantities are the inputs to the thermally activated rate expression of Eq.~\ref{eq:rate_i} and, through a Poisson event-count model, determine the conditional mechanism probability $P(\mathcal{O}_i|\mathcal{S}_t,\mathcal{M})$.
Extracting these energy barriers requires simulating rare, thermally activated transitions that constitute each mechanism, such as grain-boundary migration steps, crack-tip dislocation emission and nanovoid nucleation, sampled across a systematically varied set of microstructural configurations.

Standard molecular dynamics, even on exascale platforms~\cite{johansson2025lammps}, cannot by itself access these events at the relevant fatigue timescales.
The central limitation is the well-known mismatch between atomistic simulation timescales, which typically reach nanoseconds to microseconds, and experimental fatigue lifetimes, which extend over millions to billions of cycles.
AMD methods overcome this gap by modifying the potential energy surface so as to enhance rare-transition sampling without changing the identity or outcome of the underlying events~\cite{plimpton2020parallel, zotov2022entropy, kastner2011umbrella, zamora2016modern, voter1998accelerating}.
Bond-boosting, parallel replica dynamics, umbrella sampling, and elevated-temperature boosting, each target a different aspect of the rare-event problem, and together they can reconstruct free-energy profiles along prescribed mechanism pathways, including low-probability reverse transitions relevant to crack arrest and self-healing.
A notable recent advance is the use of machine-learning biasing potentials adapted in real time to transition states encountered during ongoing simulations \cite{musil2021physics,goff2024permutation,goff2024generalized}, which guide AMD trajectories toward previously unsampled mechanism branches and enable discovery of rare pathways inaccessible by conventional MD.
Unlike machine-learning interatomic potentials aimed at reproducing the full potential energy surface, these biasing potentials focus specifically on increasing energy barriers at previously sampled transition states, enriching the diversity of observed mechanism events without sacrificing the physical fidelity of the underlying interatomic model~\cite{wood2019data, bartok2018machine, goff2024generalized}.
Systematic coverage of the design space requires large simulation ensembles spanning grain-boundary types, triple-junction geometries, grain sizes in the range 10--100~nm, crack configurations, stress amplitudes, and temperatures.
Such combinatorial exploration is only practical through automated high-throughput workflows distributed across exascale resources~\cite{hudson2022libensemble,hudson2023libensemble}.
The resulting output is a high-dimensional dataset of labeled mechanism events $\mathcal{O}_i$ (crack-tip dislocation emissions, grain-boundary migration steps, nanovoid nucleation events, cold-welding contacts) from which $P(\mathcal{O}_i \mid \mathcal{S})$ can be estimated and its sensitivity to state variables quantified. 

\subsection{Mesoscale layer: collective energy energies and state transition probabilities}\label{sec:mesolayer}
Atomistic simulations resolve individual transition pathways and local energy barriers, but they operate on volumes and timescales too small to capture the collective dislocation and grain‑boundary ensembles that govern fatigue at the microstructural scale.
The mesoscale layer is therefore needed to populate the state‑transition distribution $P(\mathcal{S}_{t+\Delta t} \mid \{ \mathcal{O}_i \}, \mathcal{S}_t, \mathcal{M})$, and to describe how dislocation activity, grain-boundary evolution, and crack-front motion interact across hundreds of nanometers to micrometers, where the relevant state transitions are no longer governed by isolated events but by coupled sequences of events.
In this regime, phase-field~\cite{miehe2010phase,miehe2010thermodynamically, duda2015phase,miehe2016phase,bieberdorf2023grain2} and dislocation-based simulations~\cite{bertin2018fft, kohnert2021spectral, bamney2023assessing} provide complementary, physically interpretable representations: the phase‑field captures evolving morphology (grain boundaries, crack fronts, branching and arrest) while discrete dislocation dynamics (DDD) methods encode spatially resolved driving stresses and pile‑ups that bias collective transitions.

The key scientific value of this layer is that it converts local activation events into a map of collective barriers and transition pathways.
These barriers are not simple sums of atomistic contributions; rather, they reflect the topology of interacting dislocations, grain-boundary curvature, defect accumulation and stress redistribution across the microstructure.
In this sense, the mesoscale layer supplies the intermediate state space in which competing pathways are organized into super-basin-like regions of configurational accessibility, linking microscopic mechanism competition to the cycle-scale statistics required by the probabilistic framework introduced earlier.

This mesoscale role has long been recognized in the mechanics of crystalline solids, where the collective evolution of dislocations, interfaces and damage fields cannot be inferred from single-event atomistics alone.
Early large-scale studies~\cite{bulatov1998connecting} demonstrated that junction formation and destruction depend on collective dislocation interactions, while later work~\cite{dimiduk2006scale,zaiser2006scale} showed that dislocation avalanches, strain bursts and intermittent energy release are intrinsic mesoscale signatures of plasticity.
These features matter for fatigue because they directly influence void nucleation, crack advance and arrest, especially in nanocrystalline materials where grain boundaries act not merely as passive obstacles but as active participants in deformation and damage evolution.
Similarly, thermally activated dislocation bypass in obstacle-dense microstructures has shown that activation energies and attempt frequencies can be formulated at the mesoscale rather than extrapolated from isolated nanoscale events, which is essential when multiple mechanisms co-activate and the effective barrier is inherently collective~\cite{sobie2017scale,sobie2017modal,sobie2017thermal}.

Practically, mesoscale modeling produces collective energy energies and state-transition statistics by exploring the configurational space of defect density fields, grain-boundary networks and crack geometries with tailored sampling.
The resulting outputs are effective transition kernels and collective barrier estimates that quantify how likely different sequences of events are under a given microstructural and loading state.
Equally important are the mesoscale observables that connect simulation to experiment, and in some cases can also be computed directly from the mesoscale fields~\cite{vizoso2024dataset,vizoso2025decoding,vizoso2025machine}: crack-front morphologies, grain-boundary curvature and migration-step statistics, local strain-field intermittency measured by digital image correlation (DIC), and mesoscale measures of dislocation density and pile-up structure inferred from TEM, EBSD and automated image analysis.
These observables provide the physical signatures against which mesoscale predictions are validated and the transition structure that connects the state variables of Section~\ref{sec:const-mech-obs} to the conditional probability distributions of Section~\ref{sec:prob-framework-mech}.
In this way, the mesoscale layer is not a numerical convenience but a physically necessary stage in populating the conditional distributions that govern fatigue damage evolution.

\subsection{Scale bridging and active learning}
The statistical connection between the atomistic and mesoscale layers is not automatic.
Barriers computed for idealized nanoscale volumes do not translate trivially into collective transition rates for polycrystalline ensembles that govern engineering-scale fatigue response, and exhaustive sampling of the design space would require more simulations than any feasible campaign could complete.
From an algorithmic perspective, four strategies address this challenge:
Bayesian scale-bridging protocols~\cite{pinz2022data,enakoutsa2026hierarchical,tian2024data},
active learning loops that concentrate simulation effort where uncertainty is highest and the scientific payoff is greatest~\cite{hudson2022libensemble,hudson2023libensemble,duschatko2024uncertainty,holber2026physics},
adaptive time-scale integration through multirate solvers and event-driven coupling~\cite{hindmarsh2005sundials,gardner2022enabling,reynolds2023arkode}, and
machine-learning-based acceleration of mesoscale simulations to reduce the cost of exploring the configurational space~\cite{herman2020data,montes2022training,oommen2022learning,oommen2024rethinking,dingreville2024benchmarking, tian2024data, bertin2024learning,teichert2019machine}.

The scale-bridging protocol proceeds hierarchically.
Kinetic and thermodynamic quantities extracted from accelerated atomistic simulations (minimum-energy pathways, free-energy barriers and strain-dependent activation functionals)~\cite{teichert2019machine,teichert2020scale,shojaei2024bridging} are represented as conditional distributions over local microstructural descriptors.
These distributions provide stochastic inputs to mesoscale simulations, allowing uncertainty at the atomistic level to propagate into collective predictions at the mesoscale.
A scale-bridging protocol can then link the two scales by treating atomistic parameters as latent variables, mesoscale observables as the likelihood, and posterior inference as the mechanism by which scale-consistent transition probabilities are obtained~\cite{pinz2022data,enakoutsa2026hierarchical,tian2024data}.

A Bayesian approach becomes most powerful when coupled to active learning~\cite{hudson2022libensemble,hudson2023libensemble,duschatko2024uncertainty}.
When posterior uncertainty in the scale-bridging model exceeds a threshold that would materially affect predicted mechanism probabilities or state transitions, additional atomistic simulations are triggered on demand.
In this way, computational resources are redirected toward the regions of microstructural and loading space where the model is least certain or most sensitive to the inferred outcome.
The result is an adaptive simulation campaign rather than a fixed parameter sweep.

Such workflow is naturally implemented in heterogeneous form, with atomistic rare-event calculations, mesoscale simulations and surrogate evaluations executed asynchronously across available resources.
Machine-learning surrogates can further accelerate mesoscale exploration by emulating expensive phase-field or DDD--phase-field solves~\cite{herman2020data,zhang2021bayesian,montes2022training,oommen2022learning,zhang2023label,oommen2024rethinking,dingreville2024benchmarking}, learning reduced-order closures~\cite{tian2024data, bertin2024learning}, or providing fast approximations of energy functionals~\cite{teichert2019machine}.
In this way, each completed calculation sharpens the estimated probability landscape and guides where the next sample should be taken.

\subsection{The simulation-generated mechanism atlas}
Integrating the atomistic and mesoscale layers produces a structured, uncertainty-quantified dataset that we refer to as the mechanism atlas: a labeled record of mechanism activation events, effective barrier distributions, branching ratios and state-transition statistics spanning the relevant design space of grain size, grain-boundary character, loading amplitude and temperature.
Each entry associates a specific microstructural and loading configuration with conditional mechanism probabilities and state-transition distributions estimated from the combined atomistic and mesoscale ensembles.

This atlas is the primary scientific output of those simulation layers and the direct input to multimodal fusion algorithms described later.
It provides mechanistically resolved, labeled transition data with quantified uncertainty across a systematically sampled design space~\cite{chernatynskiy2013uncertainty}; information that experiments cannot supply in complete form because they typically integrate over many unresolved events. Conversely, experiments provide chemical heterogeneity, environmental complexity and statistical realism that no idealized simulation volume can fully reproduce~\cite{mcdowell2007simulation,dingreville2016review}. The fusion of these two complementary projections of the same underlying mechanistic reality is therefore essential to the broader probabilistic framework~\cite{trask2024unsupervised,walker2025unsupervised}.

Beyond being a database, the atlas functions as an organized map of the mechanism space~\cite{curtarolo2013high}.
It supports downstream tasks such as clustering of mechanistically similar states or the
identification of blind spots in the sampled design space~\cite{rosenbrock2017discovering,swinburne2023coarse}, and
retrieval of representative transition pathways for comparison with experiment. 
Its value lies not only in cataloging what mechanisms occur, but in quantifying when, where and with what uncertainty they occur across the ranges of microstructural state and loading conditions most relevant to fatigue.

\section{Fused Experimental--Simulation Signatures}
\label{sec:exp-sim-sig}

The accelerated multiscale simulation pipeline described in Section~\ref{sec:acl-multiscale-sim} produces a mechanistically labeled dataset: conditional mechanism probabilities, barrier distributions and state-transition statistics across a systematically sampled design space.
Experiments provide a complementary but structurally different dataset: partial observables that integrate over many unresolved mechanism events.
Neither alone closes the inference problem.
Simulations operate in idealized geometries stripped of chemical heterogeneity and environmental complexity,
whereas experiments cannot resolve individual activation events or assign them unambiguously to specific microstructural configurations.
What is needed is the inverse of both projections simultaneously:
inferring the full joint distribution over mechanisms and observables from two partial and structurally different views of the same underlying mechanistic reality.
This is the problem of multimodal data fusion~\cite{ramesh2022hierarchical, ock2024unimat, trask2024unsupervised, walker2025unsupervised}, and it is the central methodological challenge of this section.

The experimental suite described in Section~\ref{sec:const-mech-obs} provides complementary but individually partial projections of the mechanism space, with each modality capturing a different facet of damage accumulation.
For example, DIC-measured strain localization has been shown to predict fatigue crack initiation sites at specific grain boundaries with a mechanistic specificity not achievable by conventional surface inspection~\cite{stinville2022origins}.
A primary objective of data fusion is to identify which observable signatures, or multimodal combinations of signatures, serve as the most robust predictors of emergent outcomes.
We include SHG optical microscopy as a candidate surrogate probe~\cite{yokota2012optical, shafiei2021detection, hristu2014nonlinear, bozhevolnyi2003direct, prylepa2018material, rellaford2021characterization, smolyaninov1997near}: whether its sensitivity to symmetry-broken interfacial environments translates into mechanistically distinct, separable polarization signatures in cyclically loaded nanocrystalline metals is an open question the data fusion pipeline is specifically designed to answer, by correlating SHG signal evolution with mechanism-resolved simulation data.

\begin{figure}[htbp]
  \centering
  \includegraphics[width=0.99\textwidth]{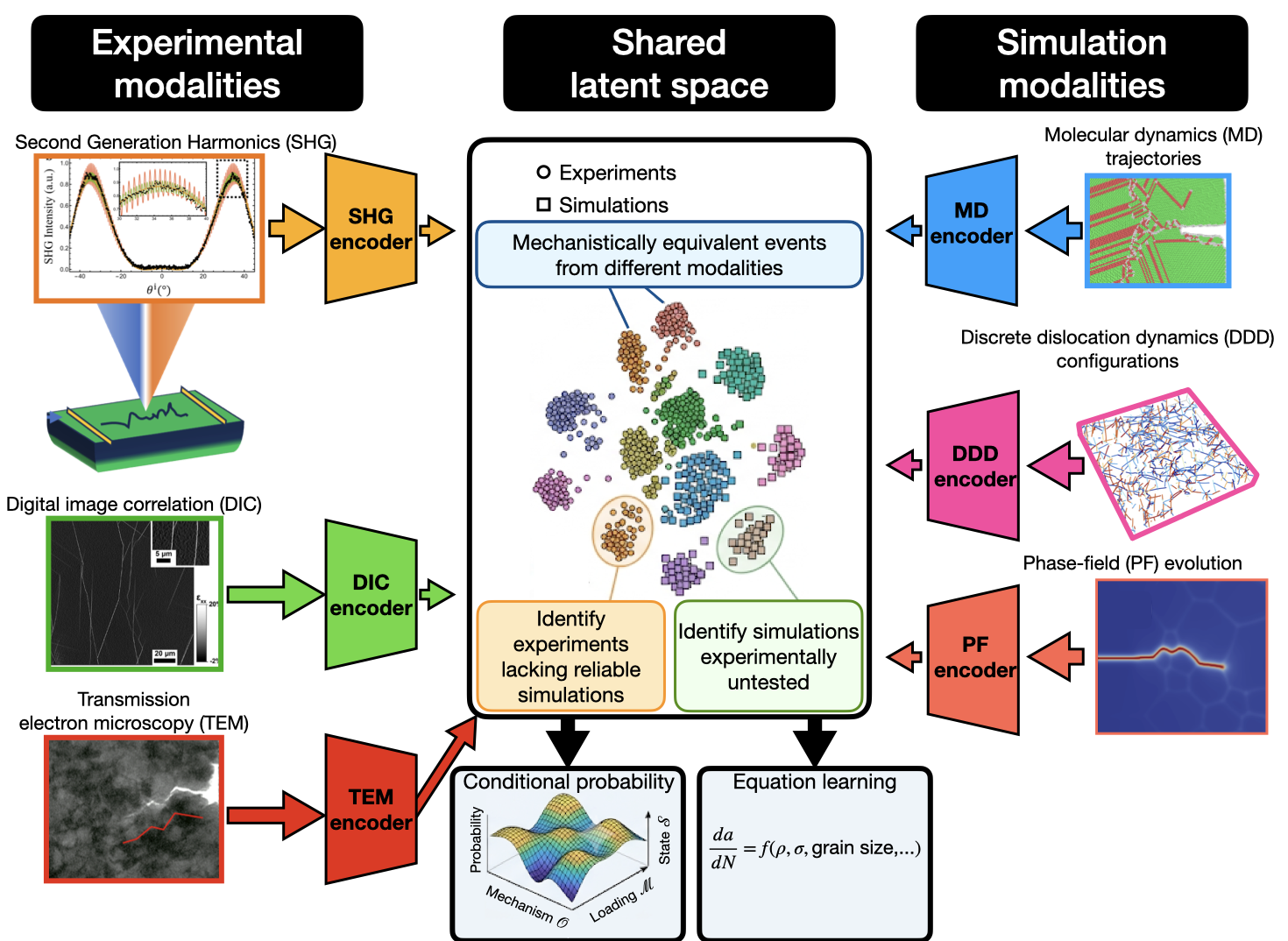}
  \caption{{\bf Multimodal data fusion architecture for inferring the conditional probability landscape from heterogeneous simulation and experimental data.}
  Two complementary data streams:
  (left) experimental modalities (in-situ TEM, DIC, SHG), each projected through a modality-specific encoder, and (right) simulation outputs (MD trajectories, DDD configurations, phase-field fields), similarly encoded. They converge into a shared latent space where mechanistically equivalent events cluster together regardless of source modality. The latent space scatter plot exposes both experiment-dense/simulation-sparse and simulation-dense/experiment-sparse regions, revealing gaps in the current mechanistic description that guide targeted simulation and experimental campaigns. Two outputs exit the shared representation: equation learning and symbolic regression for interpretable physical law discovery, and the conditional probability landscape $P({\mathcal{O}_i} | \mathcal{S}, \mathcal{M})$ that feeds directly into the Bayesian optimization and closed-loop control scheme of Section~\ref{sec:elicit}.
}
  \label{fig:datafusion}
\end{figure}

The simulation data generated by the multiscale campaign provide the mechanistically labeled counterpart to this experimental suite. The challenge is representational: the raw outputs of MD simulations, DDD configurations and phase-field fields live in spaces of such different dimensionality, geometry and physical semantics that no direct comparison with experimental signals such as SHG microscopy is possible.
A meaningful comparison therefore requires projection into a common, lower-dimensional representation space (see Fig\@.~\ref{fig:datafusion}) in which mechanistically relevant information is preserved while modality-specific artifacts and noise are discarded~\cite{rosenbrock2017discovering, trask2024unsupervised, walker2025unsupervised, robertson2021digital}.

\subsection{Shared latent space and representation learning}
\label{sec:shared-latent}

A shared latent space provides a common representational language for heterogeneous data, enabling them to be aligned without forcing them into an inappropriate common measurement basis.
The approach to such alignment rests on spatiotemporal foundation models~\cite{mccabe2024multiple, rahman2024pretraining, vaswani2017attention, liu2021swin, xu2023multimodal, oquab2023dinov2, radford2021learning, horwath2024ainerd, lipton2018mythos, murdoch2019definitions, mccabeWalrusCrossDomainFoundation2025}.
These large-scale deep learning architectures are pretrained on massive unlabeled datasets from individual experimental and simulation modalities, learning compressed latent representations that capture the dominant structure of each data stream.
For each modality, a learned encoder maps raw data into a fixed-dimensional latent vector, retaining information predictive of downstream scientific quantities (mechanism identity, damage state, fatigue life remaining...) while discarding modality-specific artifacts.
In this pipeline, patch-based representations are naturally suited to multiscale materials data: atomic neighborhoods, grain-boundary segments and grain-scale configurations can be treated as successive patches, with self-attention across levels linking local mechanism signatures to collective behavior.

\begin{figure}[htbp]
  \centering
  \includegraphics[width=0.99\textwidth]{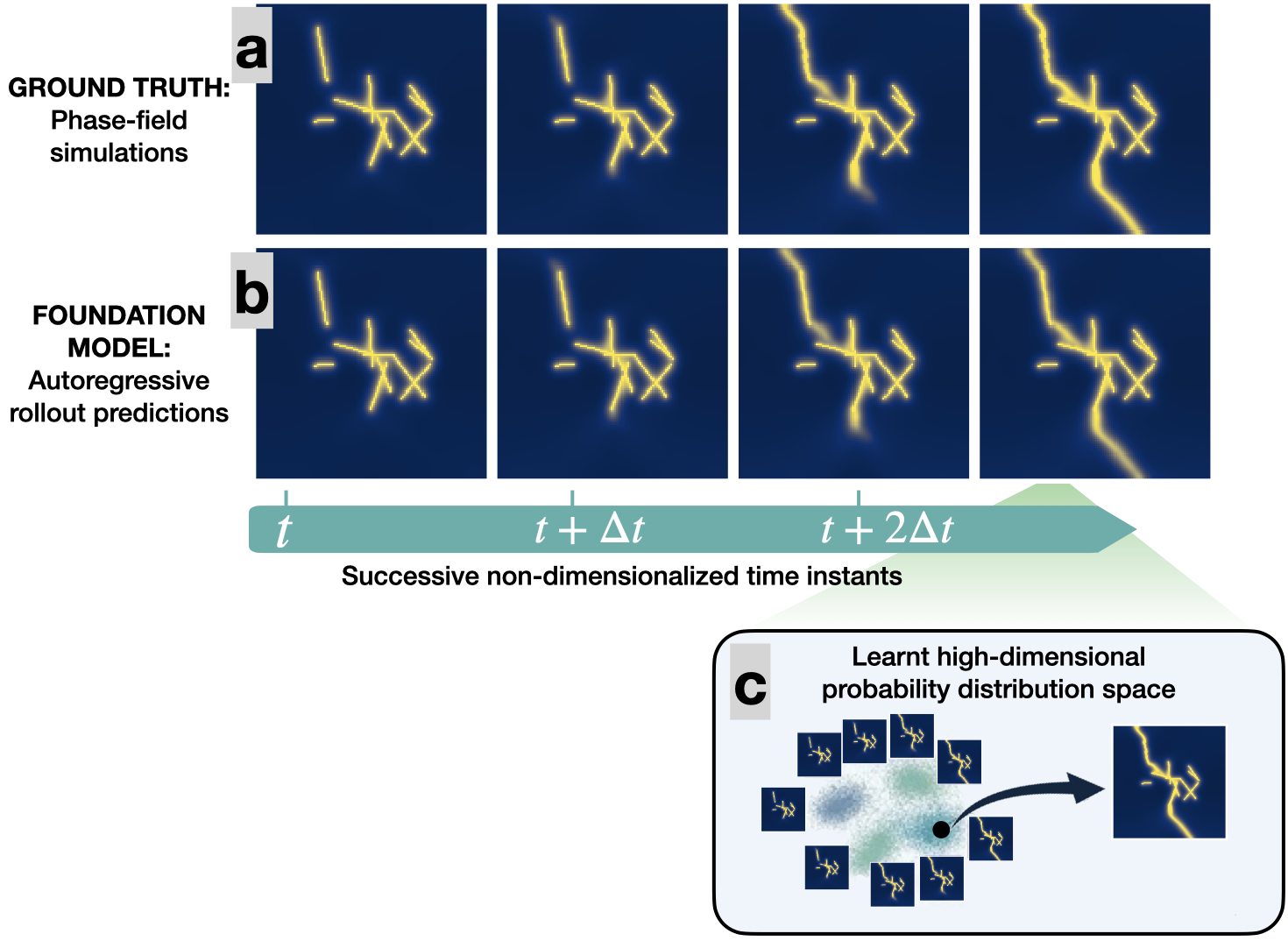}
  \caption{{\bf Foundation-model prediction of spatiotemporal fracture evolution.}
  (a) Ground-truth phase-field simulations showing the progressive evolution of a crack network in a brittle material.
  (b) Autoregressive rollout predictions from a fine-tuned Walrus spatiotemporal foundation model, forecasting crack propagation over successive non-dimensional time steps from the same initial condition. The close agreement demonstrates the model’s ability to capture the underlying fracture dynamics and reproduce mechanistically consistent evolution within a learned latent representation.
  (c) Learned high-dimensional latent probability space constructed from sequences of fracture snapshots. Each trajectory corresponds to an encoded spatiotemporal evolution path, where proximity reflects similarity in underlying fracture mechanisms and transition likelihoods. The autoregressive model advances dynamics as probabilistic transitions in this space, enabling coherent long-horizon predictions and cross-trajectory generalization.
}
  \label{fig:fondationmodel}
\end{figure}

Cross-modal fusion is accomplished by training a shared embedding layer that projects individual latent vectors into a common space, typically under contrastive constraints that place mechanistically equivalent events near one another regardless of source modality~\cite{ramesh2022hierarchical, ock2024unimat, trask2024unsupervised, walker2025unsupervised}.
The resulting shared latent space serves as a common coordinate system in which proximity reflects similarity of underlying processes and trajectories reflect evolution under loading.
This representation is not merely descriptive but also predictive: temporal trajectories in latent space can be advanced using autoregressive foundation models to forecast system evolution.
For example, Fig.~\ref{fig:fondationmodel} illustrates a fine-tuned spatiotemporal foundation model~\cite{mccabe2023multiple,mccabe2024multiple,mccabeWalrusCrossDomainFoundation2025} that accurately reproduces the progressive evolution of fracture networks, while implicitly operating in a learned high-dimensional latent probability space constructed from sequences of microstructural snapshots.
Its value lies not in replacing the original observables, but in providing a compact representation in which otherwise incommensurate data become comparable.

Complementary to foundation models at the continuum scale, machine-learning interatomic potentials (MLIAPs)~\cite{batzner20223, musaelian2023learning, Batatia2022mace} inject foundation-model capabilities into atomistic simulation.
Rather than serving as representation models for multimodal fusion, MLIAPs provide a transferable, first-principles-informed simulation layer for generating mechanism-rich trajectories across complex local environments such as crack tips, grain boundaries, nanovoids and cyclically driven configurations~\cite{tan2026high, kozinsky2023scaling}.
In this capacity, foundation MLIAPs broaden the sampling of unit mechanisms, including rare or reverse pathways relevant to crack arrest and self-healing, and thereby enrich the latent space with physically informative variation.

\subsection{Inference, equation learning, and retrieval-augmented interpretation}
\label{sec:inference-eq-learning}

Once a shared latent space has been established, it becomes possible to ask questions that are difficult to pose in either experiment or simulation alone.
The first is one of structure: which observations cluster together across modalities, and which regions of the latent space expose gaps in the sampling of mechanism space?
Unsupervised clustering can reveal experiment-dense but simulation-sparse regions that point to physically relevant processes still missing from the computational campaign, as well as simulation-dense but experiment-sparse regions that identify kinetically accessible but experimentally rare mechanisms (see Fig\@.~\ref{fig:datafusion}).
In this sense, latent-space analysis does not merely organize the data; it reveals where the current description of the probability landscape remains incomplete.

The second question is one of interpretation.
Equation-learning methods provide a route from latent patterns to symbolic relations between mechanism activation probabilities and microstructural or loading variables~\cite{desai2021parsimonious, desai2025self,desai2025autoscilab,desai2026learning}.
This is where latent-space analysis connects most directly with the Interpret stage of the DLIE cycle~\cite{tsao2024ai}: statistical regularities are distilled into candidate physical equations that can be tested, refined or rejected.
Related dynamical models, including neural differential equations~\cite{andrejevic2024data, wang2019variational, wang2021variational} extend this logic to the evolution of mechanism states, while probabilistic graphical learning~\cite{walker2024flow} can help infer conditional dependencies among variables in the fused representation.
In data-rich regions, flexible surrogate models and conditional flow matching~\cite{generale2024conditional, tong2023simulation, albergo2022building, albergo2023stochastic, tong2023improving}
can represent complex distributions, whereas Gaussian-process-based methods remain valuable where events are rare and uncertainty must be quantified explicitly~\cite{parra2017spectral, hensman2015scalable}.

A further layer of interpretation comes from retrieval-augmented generation (RAG)~\cite{lewisRetrievalAugmentedGenerationKnowledgeIntensive2021, shojaei2025aiuniversity}.
In a materials-science setting, retrieval can ground latent clusters in curated mechanistic descriptions drawn from the literature and associated databases, turning compact representations into traceable natural-language explanations.
The value of such methods is not that they automate scientific judgment, but that they preserve provenance while helping connect latent structure to established physical vocabulary. Used in this way, retrieval becomes an interpretive aid rather than a substitute for mechanistic reasoning.

Taken together, these inference tools compress the shared latent space into a practical representation of the conditional probability landscape. That landscape describes how mechanisms and observables relate, but it does not yet prescribe how to steer the system toward a desired outcome; that step is the optimization problem addressed in Section~\ref{sec:elicit}.

\section{From Prediction to Control: Eliciting Exceptional Properties \& Emergent Behaviors}
\label{sec:elicit}
The probabilistic framework of Section~\ref{sec:prob-framework-mech} recasts materials behavior as inference over conditional distributions, that is learning the conditional probability distributions $P(\{\mathcal{O}_i\} \mid \mathcal{S}, \mathcal{M})$ and $P(E \mid \{\mathcal{O}_i\}, \mathcal{S}, \mathcal{M})$ from fused simulation and experimental data.
The previous sections show how those distributions can be populated and interpreted.
The challenge addressed here is different: once the probability landscape has been inferred, how can it be navigated and ultimately exploited to make a desired emergent outcome more probable?
This is the Elicit stage of the DLIE cycle, in which the emphasis shifts from explanation to prescription.
Bayesian optimization provides the navigation strategy across the design space; closed-loop decision-making such as agentic AI connects observation to simulation and model updating, and direct microstructural engineering modifies the underlying probability landscape itself.
Once the conditional probability landscape $P(E \mid \{\mathcal{O}_i\}, \mathcal{S}, \mathcal{M})$ has been learned with sufficient fidelity, it becomes an objective function optimizable by Bayesian methods that have driven advances in drug discovery~\cite{sadybekov2023computational}, catalyst design, and autonomous laboratory systems~\cite{vriza2026operating, prince2024opportunities, schmidgall2025agent, gottweis2025towards}.
The intellectual shift is from asking what probability a desired outcome has under a given condition, to asking which conditions maximize that probability. For fatigue, that shift is especially consequential because the relevant outcomes, such as crack arrest or self-healing, may occupy narrow and highly structured regions of mechanism space that would be missed by conventional design rules.

\subsection{Bayesian optimization across fidelities and objectives}
\label{sec:bayesian-opt}
Bayesian optimization~\cite{shahriari2016taking, greenhill2020bayesian} provides the workhorse strategy for navigating the learned probability landscape.
A surrogate model, often a Gaussian process, approximates the target conditional probability as a function of design variables accessible to microstructural engineering, including grain size distribution, grain-boundary character distribution, texture, loading amplitude, load ratio and temperature.
An acquisition function then identifies the next conditions to evaluate by balancing predicted performance against uncertainty.

In practice, the design space is populated by information sources of very different fidelity and cost.
Multi-fidelity optimization~\cite{huang2006sequential,lam2015multifidelity,kandasamy2016multi,poloczek2017multi,klein2017fast,folch2023combining} offers a principled way to combine them, using inexpensive approximations to explore broadly and expensive high-fidelity calculations to validate promising regions.
Graph-based variants~\cite{gorodetsky2021mfnets,jakeman2023pyapprox,zeng2025boosting} are particularly useful when MD, DDD and phase-field calculations are not simply nested resolutions of one another, but complementary representations of different sectors of the mechanism space.
For problems with multiple competing objectives, multi-task Bayesian optimization~\cite{swersky2013multi} and composite objective functions~\cite{larson2016stochastic, ferranpousa2023bayesian,ibcdfo} can capture trade-offs between, for example, promoting healing while suppressing crack nucleation.

\begin{figure}[htbp]
  \centering \includegraphics[width=0.99\textwidth]{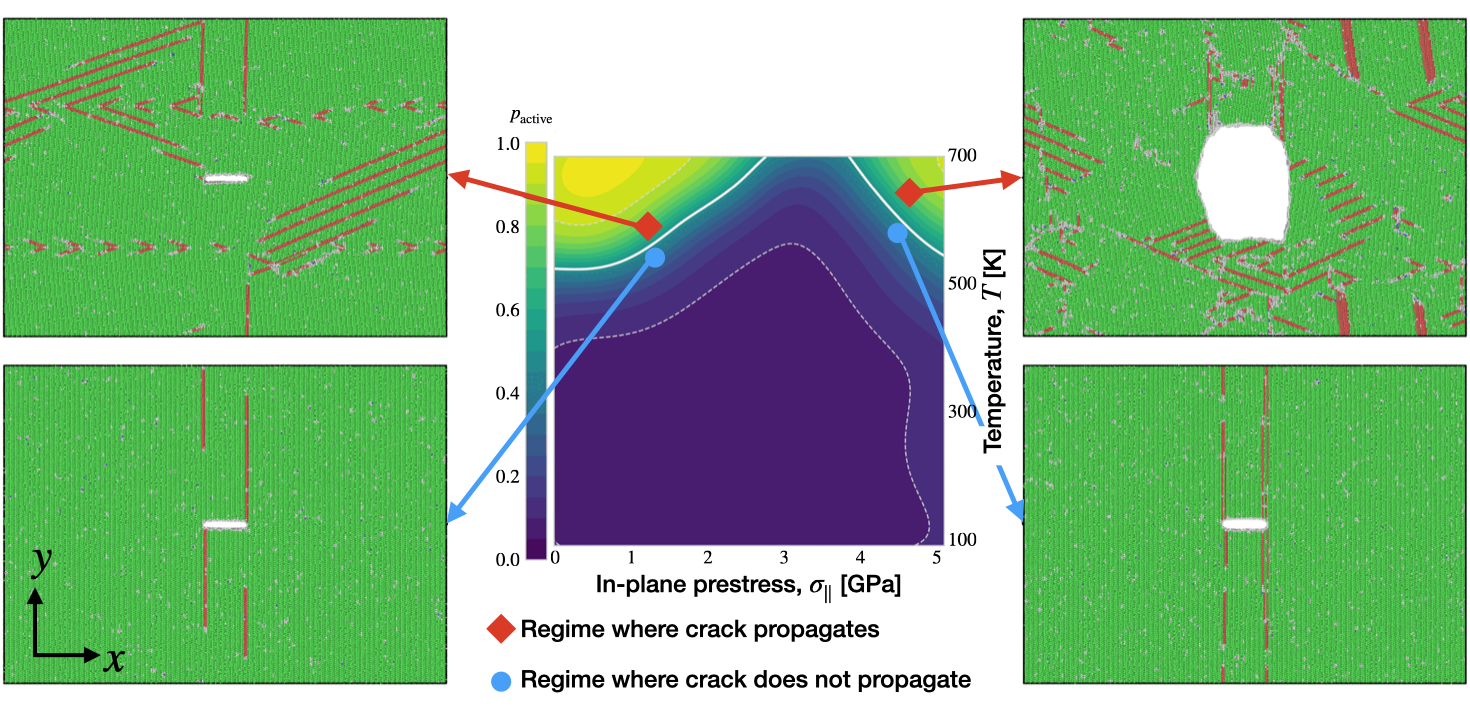}
  \caption{{\bf Bayesian level-set estimation as an edge-detection strategy for crack-growth control.} 
  A Griffith-like crack loaded in mode I (tension normal to the crack plane in the $y$ direction, $\sigma_{\perp}$) is interrogated over a two-parameter design space spanned by temperature $T$ and an in-plane prestress $\sigma_{\parallel}$ applied parallel to the crack plane ($x$ direction).
  A Gaussian-process surrogate model is used to represent the degree of crack growth under the explored conditions. In each iteration, additional points are chosen to refine the surrogate and zoom in on the boundary between active crack-growth and inactive regimes in a reduced design space. Additional simulations in the vicinity of the transition region indicate the different atomistic mechanisms at play.}
  \label{fig:BayesOptSims}
\end{figure}

As a proof of concept for such an optimization strategy, Fig.~\ref{fig:BayesOptSims} illustrates a Bayesian level-set estimation applied to the transition between crack propagation and crack arrest for a Griffith-like crack~\cite{dingreville2017primer}.
The crack is loaded in tension normal to the crack plane ($\sigma_{\perp}$, mode I), and the optimization design space is spanned by temperature $T$ and an in-plane prestress $\sigma_{\parallel}$ applied parallel to the crack plane (i.e., $x$ direction).
In this setting, the Gaussian-process surrogate progressively refines the boundary separating active and inactive crack-growth regimes, while additional simulations near the transition region reveal the atomistic mechanisms associated with each side of the boundary.
The resulting level-set estimate serves as an edge-detection strategy for identifying the onset of crack growth and, more broadly, for localizing mechanistically distinct regions of the design space.

The broader point is that optimization in this context is not a black-box search for a single best condition.
It is a structured attempt to locate the regions of mechanism space where a desired emergent behavior is robust to uncertainty, variability and competing pathways.

\subsection{Closed-loop control}\label{sec:agentic-ai}
Closed-loop control is the natural extension of optimization when the relevant mechanism landscape evolves in time, as it does under cyclic loading, evolving damage, and microstructure evolution.
In that setting, the design problem is not static: each new experiment or simulation can change the inferred probability landscape, and the control strategy must adapt accordingly.
Recent advances in agentic workflows~\cite{prince2024opportunities,schmidgall2025agent,gottweis2025towards} enable exactly this: an orchestration layer that links observation, simulation and model updating into a single adaptive cycle.

\begin{figure}[htbp]
  \centering \includegraphics[width=0.99\textwidth]{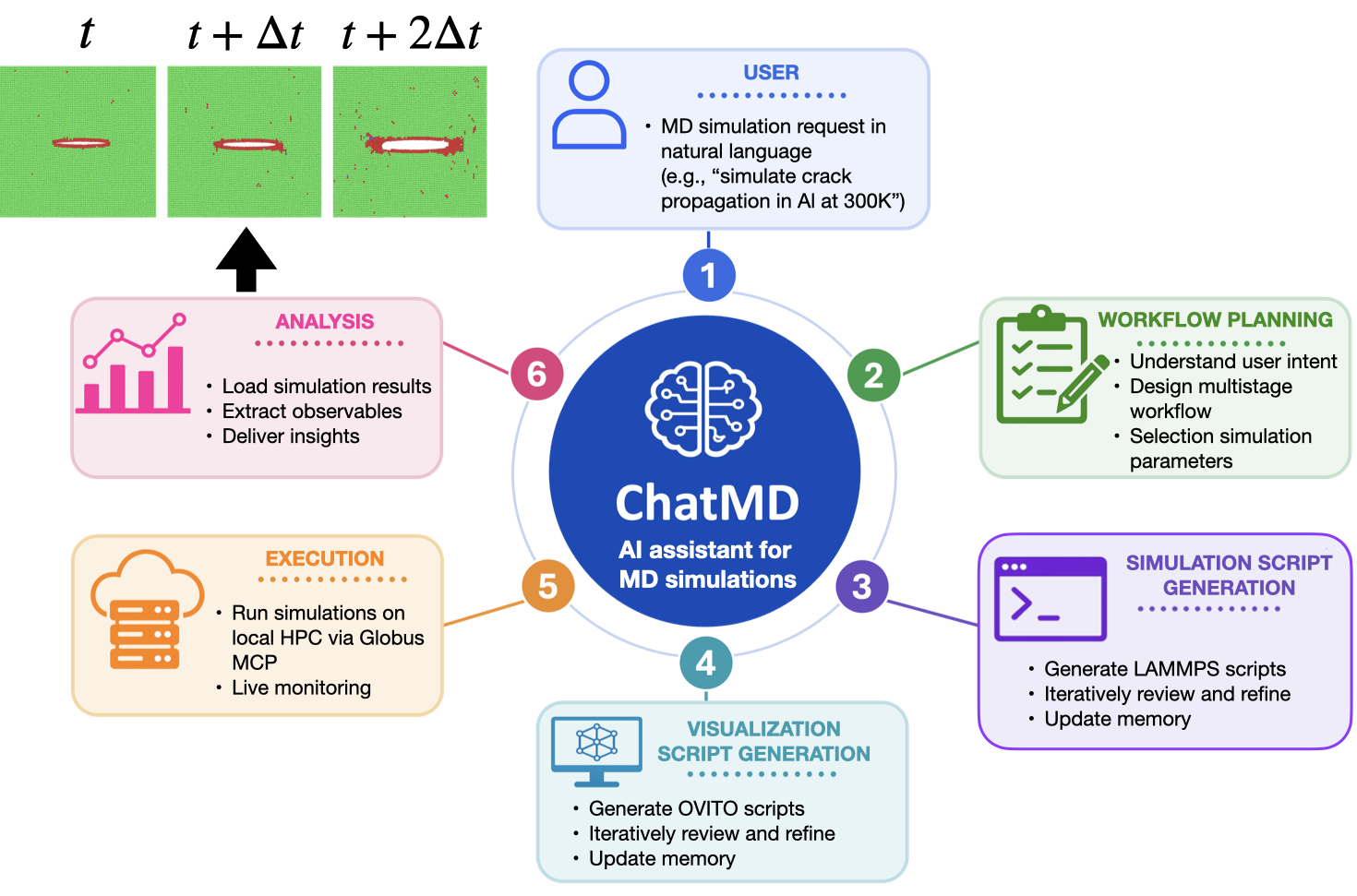}
  \caption{{\bf Multi-agent architecture  to automate and accelerate molecular dynamics simulations.} The workflow is organized into six broad phases:
  (1) natural-language user request;
  (2) workflow planning, including intent interpretation, multi-stage workflow design, and recommendation of appropriate simulation parameters;
  (3) simulation script generation through a generation-review cycle;
  (4) visualization script generation through a similar generation-review cycle;
  (5) execution on local or high-performance computing (HPC) resources; and
  (6) analysis, including extraction of physical observables, contextual question answering, and generation of scientific insights.
}
  \label{fig:agentic-workflow}
\end{figure}

Figure~\ref{fig:agentic-workflow} illustrates such orchestration for molecular dynamics, where the workflow executes multiple distinct tasks, including request interpretation, workflow planning, script generation, execution, visualization and analysis.
Rather than treating these steps as isolated manual operations, a multi-agent system can distribute them across specialized components coordinated by an large language model (LLM)-based orchestrator.
In the example in Fig\@.~\ref{fig:agentic-workflow}, natural-language requests (e.g., ``simulate the propagation of a crack in Al at 300 K'') are translated into structured simulation workflows, parameters are proposed from a knowledge base, and executable LAMMPS scripts are generated through a generation-review cycle.
Visualization can be handled in the same way, execution can be dispatched to local or high-performance computing resources~\cite{chard2017globus}, and downstream analysis can summarize observables and trends from the resulting trajectories. 
The point of such a design is not merely automation for its own sake, but the creation of an adaptive loop in which each completed simulation informs the next decision.

Incoming data, for example from DIC, AFM, in-situ TEM and SEM, supplemented where appropriate by corresponding simulation data or simulated experimental modalities can be embedded fro instance using domain-adapted vision models~\cite{zheng2024rapid} in the same representation pipeline used in Section~\ref{sec:exp-sim-sig}.
Such embeddings allow new observations to be compared directly with the current latent representation of mechanism space, highlighting regions where the probability landscape is well-constrained and regions where it remains underconstrained.
The role of automation here is not to replace scientific judgment, but to organize information flow so that simulations and measurements are directed toward the most informative gaps in the current probability landscape.
In this view, the value of closed-loop control lies in its adaptivity.
Rather than executing a fixed sequence of measurements and calculations, the workflow updates its own priorities as evidence accumulates~\cite{prince2024opportunities,schmidgall2025agent,gottweis2025towards}.
The conceptual analogue in materials science is a self-correcting map of mechanism competition: one that becomes more precise with each cycle of observation, inference and update.

\subsection{Engineering the probability landscape}
A deeper form of control goes beyond selecting favorable conditions in an existing design space and instead modifies the probability landscape itself.
Here, the objective is to alter $P(\mathcal{O}_i \mid \mathcal{S})$ through microstructural design so that the desired outcome becomes intrinsically more likely.
Grain-size gradients near crack nucleation sites, for example, can bias the competition toward grain-boundary migration and cold welding~\cite{li2020mechanical,li2024crack};
grain-boundary engineering that increases the fraction of special CSL boundaries, particularly \(\Sigma3\) coherent twins~\cite{qiu2024grain}, can suppress intergranular decohesion while maintaining boundary mobility;
and residual stress engineering can promote the compressive excursions required for crack arrest and cold welding, as observed by Barr et al.~\cite{barr2023autonomous}.

The key quantity in this setting is the learned conditional probability distribution $P(\Delta a < 0 \mid \{\mathcal{O}_i\}, \mathcal{S}, \mathcal{M})$, which maps processing-accessible variables onto the probability of a desired emergent outcome.
Yet this landscape should not be thought of as static.
Under cyclic loading, the state variables evolve, the relevant microstructural descriptors shift, and the mechanism competition itself changes.
The most realistic control strategies must therefore be interpreted as operating on a moving target, one that is continually updated by damage, healing, and further microstructural evolution.

The broader implication is that exceptional properties are not simply discovered; they are navigated toward, and in some cases engineered into existence, by shaping the conditional probabilities of the constituent mechanisms from which they emerge.
The self-healing crack reported by Barr et al.~\cite{barr2023autonomous} illustrates that the landscape of mechanism competition contains attractors absent from classical damage-tolerance thinking. 
Identifying and stabilizing such attractors is the conceptual end point of the Elicit stage.

\section{Broader Implications: Beyond Fatigue}
{
The value of the probabilistic framework developed for fatigue is not confined to crack propagation.
The broader scientific gap it addresses is recurring across many physical and chemical systems: emergent outcomes are often governed by the competition and conditional dependencies of multiple unit processes, yet standard phenomenological laws reduce that diversity to mean behavior and therefore miss the tails of the distribution where rare but consequential events occur.
In each case, the specific mechanisms differ, as do the experimental characterizations, simulation techniques, and the relevant conditioning variables, but the underlying logic is the same.
Several examples below from different scientific domains illustrate this point.
}

{
A close structural materials case may be radiation damage.
Under neutron or ion irradiation, displacement cascades produce spatially correlated Frenkel pair defect populations whose long-term evolution is governed by an ensemble of individually well-characterized constituent mechanisms including
point defect migration,
void and dislocation loop nucleation,
radiation-induced solute segregation,
helium bubble formation, and
sink-source interactions at grain boundaries and dislocations~\cite{zhang2018radiation}.
The emergent observable consequences, such as void swelling, radiation hardening, irradiation-assisted stress corrosion cracking, and helium embrittlement, are not attributable to any single mechanism in isolation.
Classical rate theory captures stochasticity at the level of individual species~\cite{dunn2016synchronous}, but it remains blind to conditional dependencies among co-active mechanisms, such as cascade-induced solute redistribution altering later defect fluxes or radiation-driven void formation modifying intergranular decohesion under stress.
A probabilistic framework over mechanism ensembles would recast these effects as a conditional probability landscape for defect evolution, naturally linking accelerated cascade simulations, mesoscale rate-theory models, multimodal characterization and Bayesian microstructure design.
}

{
In chemistry, heterogeneous catalysis illustrates the an analogous phenomenological compression~\cite{cui2022heterogeneous}.
Measured turnover frequencies and product selectivities are ensemble averages over competing adsorption, surface-reaction, and desorption pathways whose relative probabilities depend on surface topology, adsorbate coverage and thermal history~\cite{schauermann2015model,kiani2024practical}.
Legacy models such as the Br{\o}nsted-Evans-Polanyi scaling relations compress such mechanistic diversity into static energy-based descriptors (adsorption energies and activation-enthalpy proxies)~\cite{bligaard2004bronsted, greeley2016theoretical}.
While highly reliable for steady-state catalyst screening, these descriptors remain fundamentally blind to transient operating regimes where the catalyst surface state departs from ideal thermodynamic expectations.
Under severe reaction conditions, selectivity and deactivation become tail phenomena within a conditional probability distribution over surface mechanisms.
For example, the sudden self-poisoning of a Pt(110) catalyst during CO oxidation illustrates such a tail phenomenon~\cite{dey2020property,zhao2026research}: a rare fluctuation in CO coverage triggers a surface reconstruction that suppresses O$_2$ dissociation.
A probability framework over these surface mechanism ensembles, populated by \textit{ab initio} transition-state calculations, accelerated kinetic Monte Carlo simulations, and \textit{operando} spectroscopy can be brought together to map the microstructural and environmental conditions under which the mechanism mix shifts toward undesired pathways~\cite{pineda2022kinetic,chen2023machine,mou2023bridging,rangarajan2026learning}.
}

{
Classical fluid mechanics offers a third example in the subcritical transition to turbulence in shear flows.
Here, sustained turbulence or relaminarization, is governed by a competition between mechanisms that maintain localized turbulent structures, such as vortex stretching and energy regeneration, and those that dissipate them through viscosity~\cite{avila2023transition, yang2026discontinuous}.
The critical Reynolds number, the field's standard transition criterion, plays the same compressing role as the Paris law in fatigue: a useful average threshold that characterizes the average flow speed at which turbulence becomes self-sustaining, but that is blind to which mechanisms are active and in what combination.
What it cannot represent is that the mechanisms sustaining turbulence are not always in play: change the conditioning variables enough and the spatial energy transfer that feeds turbulent puffs shuts off, taking the self-sustaining cycle with it.
Recent experiments~\cite{yang2026discontinuous} show that this competition is conditioned by body forces and flow structure, so that the same Reynolds number can produce either coexistence between laminar and turbulent regions or abrupt collapse of the turbulent state.
A probabilistic formulation of sustained turbulence would for example treat the outcome as a conditional distribution over mechanism ensembles, with flow simulations, particle-image velocimetry, pressure sensing and Bayesian control strategies used to map and reshape that distribution.
}

Across these domains, the common scientific task is not to replace detailed physics with abstraction, but to identify the conditional dependencies that govern which mechanisms are accessible, which combinations co-occur and which rare pathways dominate the tails of the response.
This is the broader promise of the probabilistic perspective: it offers a portable framework for systems in which emergent behavior reflects mechanism competition under changing conditions, and where the most important physics is often hidden in events too rare for mean-field descriptions to capture.

\section{Outlook and Open Challenges}
The broader significance of the proposed probabilistic framework lies in its attempt to replace static, mean-field descriptions of materials behavior with probability landscapes of mechanism competition that can be validated, updated and ultimately used for prediction and control.
Its credibility will depend on whether those landscapes can be inferred with sufficient fidelity to support prospective tests against withheld observations.

Several open challenges define the next stage of such probabilistic approach.
Identifying which mechanisms are truly correlated and which are conditionally independent given the local state is essential for both tractability and physical interpretation.
The coarse-graining operator presented in Section~\ref{sec:prob-framework-mech} that maps lower-scale data to the mechanism labels that are relevant still depends heavily on domain expertise, and automating that step remains a prerequisite for generalizing beyond anticipated mechanisms.
Multiscale bridging is also incomplete: accelerated atomistic methods and mesoscale simulation span many orders of magnitude, but a substantial gap to engineering lifetimes remains, and closing it without uncontrolled approximations is still an open challenge.
Finally, because the conditional probability landscape evolves as damage accumulates and microstructure changes, the probabilistic framework will need efficient strategies for continual updating, including extension to incorporating or eliminating mechanisms and state variables whose probabilities have changed significantly, if it is to remain predictive under realistic service conditions.

We realize that the framework at community scale will require shared infrastructure rather than isolated effort.
That means open-source toolchains that connect simulation, uncertainty quantification and workflow management into a documented and maintainable ecosystem; curated repositories with standardized schemas that store mechanistic labels and conditional probability metadata alongside conventional materials descriptors; and community benchmarks that play for mechanism-probability datasets a role analogous to that of ImageNet~\cite{russakovsky2015imagenet} or the Materials Project~\cite{horton2025accelerated} in their respective fields.
No single group can generate the experimental coverage, simulation breadth and characterization depth needed to populate and test such a framework.
The scientific return will depend on coordination, interoperability and a willingness to treat probabilistic mechanism data as a community resource.

The broader ambition of this perspective is a shift in how materials science relates to the systems it studies:
from correlating structure and performance after the fact
to identifying, in advance, the conditions that make desired emergent behavior probable.
Whether that shift is achieved will depend on whether conditional probability landscapes can be learned with sufficient fidelity to extrapolate beyond the training distribution, infer causally meaningful mechanism couplings from partial observations and remain robust as the material evolves in time.
The motivating self-healing crack exemplar reported by Barr et al.~\cite{barr2023autonomous} shows that such landscapes contain attractors that classical damage tolerance did not seek, because the conceptual probabilistic framework needed to recognize them was not yet available.
Building that framework is the central challenge, and the central opportunity, left by this perspective.

\section*{CRediT authorship contribution statement}
{\bf R.D.}: Conceptualization, Funding acquisition, Methodology, Project administration, Supervision, Writing – original draft, Writing – review \& editing.
{\bf B.L.B., B.D., S.D., P.I., M.A.W., M.C., J.L., T.M., I.S., D.J.G., L.C., K.G.}: Conceptualization, Funding acquisition, Methodology, Writing – review \& editing.
{\bf P.R., A.E.R., B.S., B.A.J., J.T., T.M., M.D.}: Writing – original draft, Writing – review \& editing.

\section*{Declaration of competing interest}
\noindent The authors declare no competing interests. 

\section*{Acknowledgements}
The authors would like to thank Agus Poerwoprajitno and Cyrus Jordan from Sandia National Laboratories for comments and suggestions on the heterogeneous catalysis and turbulence examples discussed in Section 7.
\newline

\noindent \textbf{Funding:} 
All the authors in this work are supported by the U.S\@.~Department of Energy, Office of Science, Office of Advanced Scientific Computing Research and  Office of Basic Energy Sciences, Scientific Discovery through Advanced Computing (SciDAC) program under the MIRAGE project.
Some of the capabilities and work described in this paper were developed at the Center for Integrated Nanotechnologies (CINT), an Office of Science user facility operated for the U.S\@.~Department of Energy.
This article has been authored by an employee of National Technology \& Engineering Solutions of Sandia, LLC under Contract No\@.~DE-NA0003525 with the U.S\@.~Department of Energy (DOE).
The employee owns all right, title, and interest in and to the article and is solely responsible for its contents. The United States Government retains and the publisher, by accepting the article for publication, acknowledges that the United States Government retains a non-exclusive, paid-up, irrevocable, world-wide license to publish or reproduce the published form of this article or allow others to do so, for United States Government purposes.
The DOE will provide public access to these results of federally sponsored research in accordance with the DOE Public Access Plan {https://www.energy.gov/downloads/doepublic-access-plan.}

\section*{Data Availability}
Data sharing not applicable to this article as no datasets were generated or analyzed during the current study.
The code and data used to support the findings illustrated in Figures 1, 4, 5, 6, 7, and 8 are available in the public repository on GitHub at \url{https://github.com/mirage-scidac}. 

\bibliography{mirage-refs}

\end{document}

%% file: nomenclature.tex
\begin{tcolorbox}[
  colback=white, colframe=black!75, boxrule=0.5pt, arc=2pt,
  left=5pt, right=5pt, top=4pt, bottom=4pt, boxsep=2pt,
  title=\textbf{Nomenclature}, fonttitle=\small,
  coltitle=black, colbacktitle=black!8]
\footnotesize
\setlength{\tabcolsep}{2pt}
\renewcommand{\arraystretch}{1.05}
\noindent
\begin{minipage}[t]{0.49\linewidth}
\begin{tabular}{@{}>{\raggedright\arraybackslash}p{0.30\linewidth}@{\hspace{4pt}}>{\raggedright\arraybackslash}p{0.64\linewidth}@{}}
\multicolumn{2}{@{}l}{\textit{\textbf{Probabilistic framework}}}\\
$E$ & emergent macroscopic observable (e.g.\ crack healing, damage tolerance)\\
$\mathcal{O}_i$ & $i$-th constituent (unit) mechanism at the chosen level of the hierarchy\\
$\{\mathcal{O}_i\}_{i=1}^{N}$ & set of $N$ co-active mechanisms\\
$\mathcal{S}$ & local subsystem state ($\mathcal{S}_t$ at step $t$; $\mathcal{S}_{t+\Delta t}$ at subsequent step)\\
$\mathcal{S}_{0:\tilde{t}}$ & time sequence of states over instants $\{0,\ldots,\tilde{t}\}$\\
$\mathcal{M}$ & global conditioning variables (microstructure, loading, environment)\\
$\lambda_i$ & expected event count for $\mathcal{O}_i$ in $\Delta t$, $\lambda_i\!=\!\dot{p}_{\mathcal{O}_i}\Delta t$ (Poisson parameter)\\
$c_i$ & coarse-graining operator; $\mathcal{O}_i\!=\!c_i(s_{t:t+\Delta t},\mathcal{S}_t,\mathcal{M})$ maps lower-scale states to occurrences of $\mathcal{O}_i$\\
$s_{t:t+\Delta t}$ & lower-length-scale state trajectory over $[t,\,t+\Delta t]$\\
$P(\{\mathcal{O}_i\}_{i=1}^N\!\mid\!\mathcal{S}_t,\mathcal{M})$ & joint probability that a mechanism set co-occurs given local state and global variables\\
$P(\mathcal{S}_{t+\Delta t}\!\mid\!\{\mathcal{O}_i\},\mathcal{S}_t,\mathcal{M})$ & state-transition distribution conditioned on activated mechanisms\\
$P(E\!\mid\!\mathcal{S}_{0:\tilde{t}},\mathcal{M}_{0:\tilde{t}})$ & probability of macroscopic observable conditioned on time sequences of states and global variables\\[2pt]
\multicolumn{2}{@{}l}{\textit{\textbf{Constituent mechanisms (crack-tip)}}}\\
$\mathcal{O}_1$ & crack-tip dislocation emission (forward slip, primary system)\\
$\mathcal{O}_2$ & crack-tip dislocation absorption (secondary slip, non-coplanar)\\
$\mathcal{O}_3$ & point defect stress-assisted diffusion\\[2pt]
\multicolumn{2}{@{}l}{\textit{\textbf{State variables}}}\\
$\rho_m$ & mobile dislocation density at the crack tip\\
$\rho_f$ & forest (immobile) dislocation density at the crack tip\\
$d_v$ & nanovoid density\\
$a$ & crack length\\
$\Delta a$ & crack-length increment per step; $\Delta a\!=\!a_{t+\Delta t}\!-\!a_t$ ($\Delta a<0$: healing)\\
\end{tabular}
\end{minipage}\hfill
\begin{minipage}[t]{0.49\linewidth}
\begin{tabular}{@{}>{\raggedright\arraybackslash}p{0.32\linewidth}@{\hspace{4pt}}>{\raggedright\arraybackslash}p{0.62\linewidth}@{}}
\multicolumn{2}{@{}l}{\textit{\textbf{Loading variables}}}\\
$R$ & load ratio, $\sigma_{\min}/\sigma_{\max}$, $\sigma_{\min},\sigma_{\max}$ cyclic extrema\\
$T$ & temperature; $k_B$ Boltzmann constant\\
$\sigma_{\perp}$ & loading stress normal to crack plane (cyclic loading condition in Bayesian optimization)\\
$\sigma_{\parallel}$ & in-plane prestress (design variable in Bayesian optimization)\\[2pt]
\multicolumn{2}{@{}l}{\textit{\textbf{Kinetics and thermodynamics}}}\\
$\dot{p}_{\mathcal{O}_i}$ & rate of mechanism $\mathcal{O}_i$ (events per unit time)\\
$\nu_{0,i}$ & attempt frequency of mechanism $i$\\
$N_i(\mathcal{S})$ & number of equivalent sites/events for mechanism $i$ in state $\mathcal{S}$\\
$\Delta G_i$ & Gibbs free-energy barrier; $\Delta G_i\!=\!\Delta H_i\!-\!T\Delta\mathbf{s}$\\
$\Delta H_i$ & enthalpy (activation) barrier of mechanism $i$\\
$\Delta\mathbf{s}$ & activation entropy (configurational + vibrational contributions)\\
$\Delta t$ & time-step: long relative to individual events; short relative to macroscopic damage accumulation\\[2pt]
\multicolumn{2}{@{}l}{\textit{\textbf{Macroscopic observables \& Paris law}}}\\
$K$ & stress intensity factor; $\Delta K$ its cyclic range\\
$\mathrm{d}a/\mathrm{d}N$ & fatigue crack-growth rate per cycle\\
$N$ & number of fatigue cycles; also index bound in $\{\mathcal{O}_i\}_{i=1}^N$\\
$C$,\,$m$ & Paris-law empirical constants; effective averages over mechanism and microstructural distributions at calibration\\
$\langle\Delta a/\Delta N\rangle_{\mathcal{M}}$ & expected crack-length increment marginalized over mechanism activations, states for a set of global variables\\
$P(\Delta a\!<\!0\!\mid\!\{\mathcal{O}_i\},\mathcal{S},\mathcal{M})$ & probability of crack reversal (self-healing); target quantity for Bayesian optimization in Sec.~\ref{sec:elicit}\\
\end{tabular}
\end{minipage}
\end{tcolorbox}